\documentclass[aps,prx,reprint,amsfonts,twocolumn,floatfix,hidelinks,superscriptaddress]{revtex4-2}
\usepackage[T1]{fontenc}
\usepackage{graphicx}
\usepackage{dcolumn}
\usepackage{amsmath,amssymb,amsfonts,bm,dsfont,units}
\usepackage{hyperref}
\hypersetup{colorlinks=true,citecolor=blue,linkcolor=blue,urlcolor=blue,pdfstartview=FitH}
\usepackage{braket}
\usepackage{float,latexsym}
\usepackage{multirow}
\usepackage[normalem]{ulem} 
\usepackage[caption=false]{subfig} 
\usepackage{bbm}
\usepackage{braket}
\usepackage{array}
\usepackage{xcolor}
\usepackage{graphicx}
\usepackage{hhline,booktabs}
\usepackage{makecell}
\usepackage{enumitem}

\usepackage[utf8]{inputenc}
\usepackage{array, multirow}

\newcolumntype{s}{>{\columncolor{blue!20}} c}

\usepackage{soul}
\usepackage{color}
\definecolor{dkgreen}{rgb}{0,0.5,0}
\definecolor{midnightblue}{rgb}{0.39,0.58,0.93}

\usepackage{mathtools}

\begin{document}

\title{Quantum Fisher Information under decoherence with explicit wave functions}

\author{Francesco Musso}
\email{francesco.musso@pasqal.com}
\affiliation{Universit\'e Paris-Saclay, CNRS, LPTMS, 91405, Orsay, France}
\affiliation{Pasqal SAS, 24 rue Emile Baudot, 91120 Palaiseau, France}
\author{Vittorio Vitale}
\affiliation{Pasqal SAS, 24 rue Emile Baudot, 91120 Palaiseau, France}
\author{Sara Murciano}
\affiliation{Universit\'e Paris-Saclay, CNRS, LPTMS, 91405, Orsay, France}

\date{\today}

\begin{abstract}
    We present a method to estimate the quantum Fisher information (QFI) of many-body quantum states in the presence of decoherence, where its direct evaluation requires the full spectral resolution of the density matrix. We show that, for many-body wave functions known analytically in the occupation-number basis, systematic lower bounds to the QFI can be mapped onto expectation values over a classical probability distribution defined by the wave function amplitudes. This mapping enables efficient estimation via Markov-chain Monte Carlo sampling, with a computational cost that scales as a `slow' exponential ($e^{b L}$ with $b \lesssim 0.6$) and remains manageable for system sizes well beyond exact diagonalization. We specify this framework to Jastrow–Gutzwiller wave functions. We characterize their metrological content by identifying the observables that maximize the QFI and the corresponding scaling with $L$. Then, we analyze the QFI under three physically motivated noise channels: local dephasing, local amplitude damping, and global depolarizing. We compare polynomial and Krylov-based lower bounds across these channels, relating their behavior to the effective rank of the noisy density matrix and to the structure of the operator generating the parameter encoding. The framework extends naturally to other analytically known wave functions and to a broader class of information-theoretic quantities beyond the QFI.
\end{abstract}

\maketitle

Quantum sensing enables the detection of weak signals and fields with extremely high precision, with applications ranging from atomic clocks and magnetometry to gravitational wave detection~\cite{magnetometry1,magnetometry2,LIGO,atomiclocks1}. In interferometric protocols, an unknown parameter $\theta$ is encoded in a quantum state $\rho$ through a unitary transformation generated by a Hermitian operator $O$. The ultimate achievable precision is given by the quantum Cramér–Rao bound, $\delta\theta \geq 1/\sqrt{F_Q[\rho,O]}$, where the quantum Fisher information (QFI) reads~\cite{Pezze2009,Braunstein1994}
\begin{equation}\label{eq:qfi}
    F_Q[\rho, O] = 2\sum_{\lambda_i + \lambda_j>0} 
    \frac{(\lambda_i - \lambda_j)^2}{\lambda_i + \lambda_j} 
    \left|\bra{i}O\ket{j}\right|^2,
\end{equation}
with $\rho = \sum_i \lambda_i \ket{i}\bra{i}$ the spectral decomposition of the state.
For $L$ qubits, separable (or classically correlated) states satisfy $F_Q \leq L$~\footnote{This only holds when $O$ is a collective spin operator $O=\frac{1}{2} \sum_{j = 1}^{L} \alpha_j \sigma_{j}^{(\tau)}$, where $\alpha_j$ are real numbers and $\sigma_{j}^{(\tau)}$ is the Pauli matrix in an arbitrary direction $(\tau)$ acting on the $j^{\rm th}$ spin (identity operators on the other qubits are implicit).}, defining the standard quantum limit (SQL), $\delta\theta \gtrsim 1/\sqrt{L}$, while entangled states can reach the Heisenberg limit $F_Q \sim L^2$, corresponding to $\delta\theta \sim 1/L$~\cite{Pezze2009,quantumsensing,Giovannetti2011,giovannetti2004}. Therefore, the QFI quantifies the ultimate sensitivity of a quantum probe and plays a central role in both quantum sensing and quantum information theory~\cite{pappalardi2017multipartite,silvia1,silvia2,silvia3,silvia4,Hauke2016,Toth2012,Hyllus2012,Pezze2009,Gietka2022}. 

Evaluating the QFI for many-body states under realistic noise is a hard computational problem:  Eq.~\eqref{eq:qfi} requires the full spectral resolution of $\rho$,  which is inaccessible for mixed states arising from decoherence at system sizes of metrological interest.
This challenge has sparked interest along two distinct lines of research. On the experimental side, a series of works has developed lower bounds to the QFI that are accessible via randomized measurements or shadow tomography~\cite{elben2023randomized,huang2020predicting,cerezo2021sub,goldman2021,Cerezo2022variational,zhang2016detecting}: in particular the converging polynomial bounds $F_n$ of Refs.~\cite{Rath2021,VitalePRX2024} and, more recently, the Krylov-subspace bounds $B_n$ of Refs.~\cite{Zhang2025,Wang2026} that can be estimated from a finite set of moments $\operatorname{Tr}(\rho^r O \rho^s O)$. On the theoretical side, the identification of many-body states that combine enhanced QFI scaling with robustness to noise remains largely open, with most analytical results confined to GHZ, spin-squeezed states and, more recently, critical states~\cite{Ma2011,Pezze2018,Montenegro2024,chen2021effects,ilias2022criticality,Chai_2025,Xu2023,ferro2025,Sara,Chen2026,Kurdzia2023,Kurdzia2025,Demkowicz-Dobrzanski2009,Dorner2009,Zhou2018}.

Within this framework, our contribution is twofold.
First, we show that for many-body wave functions known analytically in the occupation-number basis, the moments entering both families of QFI bounds can be recast as expectation values over a classical probability distribution defined by the wave function amplitudes, after the action of generic quantum channels. This recasting turns the evaluation of QFI bounds for noisy many-body states into a Markov-chain Monte Carlo sampling task~\cite{becca2017quantum}, with a cost scaling as $\sim e^{b L}$ with $b \lesssim 0.5$.
Second, we investigate the family of Jastrow–Gutzwiller (JG) wave functions as a metrological resource~\cite{haldane1988exact,OnuttomNarayan,becca2017quantum,Turkeshi2020,cirac09,stephane17,Herwerth18,FrancoRubio2025}. These states interpolate continuously between paradigmatic limits: on one side, highly multi-partite entangled Greenberger–Horne–Zeilinger (GHZ)-like states exhibiting Heisenberg scaling; on the other, critical states described by Luttinger liquid theory, where correlations decay algebraically. This interpolation provides a unified platform to explore different metrological regimes within a single analytic family of many-body states.

We start by characterizing the JG wave functions in the absence of noise. By identifying suitable operators $O$, we show that these states can achieve sensitivities beyond the SQL across a wide parameter range, including regimes with Heisenberg scaling. Importantly, the optimal observable depends non-trivially on the correlation structure encoded in the wavefunction.
We then apply our framework on three physically motivated noise channels: local dephasing, local amplitude damping, and global depolarizing, chosen to span qualitatively different regimes for noise channel action, density-matrix rank, and operator structure.
We find that across these three cases, the bounds exhibit qualitatively different behaviors. In agreement with the recent Ref.~\cite{Wang2026}, we find that $B_n$ provides tighter bounds than $F_{2n-1}$ at fixed moment budgets. Our analysis additionally shows that the practical gap between the two families depends sensitively on the effective rank of the noisy density matrix and the matching between operator and channel. Low-rank states yield rapid convergence, high-rank states with broadly distributed spectral weight yield systematically looser bounds, and operator structure can  improve convergence even at high rank when it restricts the transitions (the matrix elements $\bra{i}O\ket{j}$) contributing to Eq.~\eqref{eq:qfi}.

Our approach establishes a direct bridge between explicit many-body wave functions and scalable estimation of metrological quantities in noisy quantum systems. Beyond the QFI and JG, the same framework can be extended to compute a wide range of information-theoretic observables, including purities and correlation functions in mixed states, and to other analytically known wave functions such as fractional quantum Hall states~\cite{TsuiStormerGossard1982,Laughlin83}.

The work is structured as follows. In Sec.~\ref{sec:jastrows} we introduce the Jastrow-Gutzwiller wave functions and discuss the QFI in the case of a pure system. In Sec.~\ref{sec:bounds} we discuss converging lower bounds to the QFI that have been employed in the literature for mixed states. In Sec.~\ref{sec:QFIwithdecoherence} we study the QFI of noisy Jastrow-Gutzwiller wave functions, obtained by the application of different quantum channels, namely local dephasing, amplitude damping and global depolarizing. Finally, we draw our conclusions. 

In the appendices we provide more technical details. In App.~\ref{app:luttinger} we study the correlations of the JG wave function in the critical regime, while in App.~\ref{app:montecarlo} we discuss the Monte Carlo sampling method and its convergence. In App.~\ref{app:dephasing} and ~\ref{app:damping} we derive analytical expressions for the QFI.
In the case of the dephasing channel, we obtain explicit estimates in the whole phase diagram of the JG wave function, while in the case of the local amplitude damping channel we obtain an analytical expression in the limit of the JG wave function approaching a GHZ state.

\section{Quantum Fisher Information of Jastrow-Gutzwiller wave functions}\label{sec:jastrows}
In this work, we consider a periodic chain of $L$ sites with occupation numbers $n_i \in \{0,1\}$. In the Fock basis, the JG wave functions take the form~\cite{Turkeshi2020}
\begin{equation}
    \label{eq:wf}
    \begin{aligned}
        \ket{\Psi_\alpha} & =\sum_{\mathcal{P}_N\{n\}} \psi_\alpha(\{n\})\ket{n_1 n_2 \ldots n_L} \\
        \psi_\alpha(\{n\}) & =\frac{1}{Z_\alpha}(-1)^{\sum_{i=1}^L j n_j} \prod_{1 \leq i<j \leq L}^L \sin \left(\frac{\pi}{L}(j-i)\right)^{\alpha n_i n_j}
    \end{aligned}
\end{equation}
where the sum runs over all configurations $\mathcal{P}_N\{n\}$ with fixed particle number $\sum_i n_i = N$, and $Z_\alpha$ is a normalization constant. We focus on half filling, $N=L/2$ with $L$ even. The state is invariant under the $U(1)$ symmetry associated with particle-number conservation, and additionally under the particle–hole transformation generated by $\prod_j X_j$ (acting as $n_j \to 1 - n_j$).

The parameter $\alpha$ controls the effective interaction between occupied sites: when $\alpha>0$, configurations in which particles are far apart are favored, while $\alpha<0$ enhances configurations where particles cluster together. 
The structure of the wave function simplifies in two extreme limits. 
For $\alpha \gg 1$, the dominant contributions come from configurations that maximize the product of the sine factors, i.e., that keep particles as far apart as possible. At half filling, this corresponds to a perfectly alternating pattern of occupied and empty sites. There are only two such configurations, related by a one-site translation, and they are degenerate. In this regime, the JG wave function is therefore well approximated by a coherent superposition of these two configurations yielding an antiferromagnetic GHZ state
\begin{equation} \label{eq:ghz} \ket{\Psi_{\alpha }} \simeq \frac{1}{\sqrt{2}}\left((-1)^{(L/2)\ \text{mod}\ 2} \ket{101 \ldots 10}+\ket{010 \ldots 01}\right). 
\end{equation}
In contrast, for $\alpha \ll -1 $, the weight is maximized by configurations that minimize the product of sine factors, favoring particles occupying a contiguous block of $L/2$ sites. Since the block can start at any lattice position, there are $L$ such configurations, and the JG state becomes a coherent superposition of all such translated blocks, resulting in a sort of Dicke state 
\begin{equation}
\label{eq:cdicke}
    \ket{\Psi_\alpha} \simeq \frac{1}{\sqrt{L}} \sum_{i=1}^L \ket{\ldots 0_{i-1} 1_i 1_{i+1} \ldots 1_{i+L / 2-1} 0_{i+L / 2} \ldots}
\end{equation}
Away from these limits, many configurations contribute to the total wave function with comparable weight, and the competition between repulsive and clustering tendencies gives rise to a richer but less tractable structure. Ref.~\cite{cirac09} showed that, for $0<\alpha \leq 2$, the JG wave functions correspond to the critical XXZ chain. Since the latter is described at low energies by a Luttinger liquid, this implies that the JG states are likewise captured by a Luttinger-liquid description, with Luttinger parameter $K=1/\alpha$. Thus, $\alpha$ continuously controls both the strength of correlations and the decay exponents of observables. In particular, the points $\alpha=0,1,2$ are exactly solvable and correspond respectively to the ferromagnetic point, the free-fermion XX point, and the isotropic Heisenberg point with $SU(2)$ symmetry. Numerical evidence from Ref.~\cite{Turkeshi2020} suggests that this description remains valid up to $\alpha \simeq 4$. While the $X$-correlators remain in very good agreement with the expected power-law behavior, the $Z$-correlators exhibit much stronger finite-size effects, preventing a sharp and reliable extraction of the exact value of $K$. We provide a detailed analysis of the subtleties of this regime in App.~\ref{app:luttinger}.

Let us now turn to the metrological properties of these states. 
For pure states, Eq.~\eqref{eq:qfi} simplifies to $F_{Q}[\rho, O]=4\mathrm{Var}(O)$, and the QFI can be evaluated exactly for any local or few-body operator $O$.
Thus, we can scan the different values of $\alpha$ and find the operator with the maximal scaling in system size $L$ for the JG state. In the regimes of very large or negative $\alpha$, this analysis can be easily done using the simplified form of the states in Eqs.~\eqref{eq:ghz} and \eqref{eq:cdicke}, respectively. In the GHZ limit, the two-configurations in Eq.~\eqref{eq:ghz} straightforwardly yield the Heisenberg limit. The staggered magnetization $O_Z = \tfrac{1}{2}\sum_j (-1)^j Z_j$ has a macroscopic variance, $F_Q \sim L^2$, because the two superposed configurations have opposite staggered order. In the clustered limit, a similar mechanism occurs: the state in Eq.~\eqref{eq:cdicke} is a coherent superposition of macroscopically distinct magnetization patterns. An operator sensitive to this structure would yield again $F_Q \sim L^2$. 
By investigating the behavior of the correlation functions for the more general case $\alpha<0$, one can find the best operator to be
\begin{equation}\label{eq:obigstar}
    O_{\bigstar}=\frac{1}{2}\sum_{j=1}^{\lfloor L/4\rfloor} Z_j - \frac{1}{2}\sum^{\lfloor 3L/4\rfloor}_{j=\lfloor L/4\rfloor+1} Z_j + \frac{1}{2}\sum_{j=\lfloor 3L/4\rfloor+1}^L Z_j,
\end{equation}
which returns a variance $\operatorname{Var(O_{\bigstar})}\sim L^2/12$.

The most physically interesting regime is the critical phase, where the scaling of the QFI can be extracted from the asymptotic decay of correlation functions described by a Luttinger liquid~\cite{Turkeshi2020,Giamarchi2003}. We consider two staggered operators $O_{Z}=\tfrac{1}{2}\sum_j (-1)^jZ_j$ and $O_{X}=\tfrac{1}{2}\sum_j (-1)^jX_j$, whose variances are determined by the corresponding two-point functions
\begin{equation}
    \langle Z_0Z_r\rangle \sim r^{-2/\alpha}, \quad    \langle X_0X_r\rangle \sim r^{-\alpha/2}.
\end{equation}
Summing over $r$, the variance scales as $\sim L^{2-\gamma}$, where $\gamma$ is the corresponding decay exponent.
Therefore, either $O_{Z}$ or the $O_{X}$ operator shows a scaling of the variance faster than the SQL, depending on the value of $\alpha$.
In the following, the operators $O_Z$, $O_X$ and $O_\bigstar$ will be used to estimate the metrological content of JG wave functions also in the presence of noise, as a function of the parameter $\alpha$, as we summarize in Table~\ref{table:var}.

\begin{table}[t!]
\label{table:var}
\setlength{\tabcolsep}{8pt}
\begin{tabular}{|c|c|c|}
\hline
$\alpha$ & Operator $O$ & $\mathrm{Var}(O)$ \\
\hline
$>4$ & {$O_Z=$} $\tfrac{1}{2}\sum_j(-1)^j Z_j$ & $\sim L^2$ \\
\hline
$(2,4]$ & {$O_Z=$} $\tfrac{1}{2}\sum_j(-1)^j Z_j$ & $\sim L^{2-\frac{2}{\alpha}}$ \\
\hline
$[0,2]$ & $O_X=$ $\tfrac{1}{2}\sum_j(-1)^j X_j$ & $\sim L^{2-\frac{\alpha}{2}}$ \\
\hline
$<0$ & 
\makecell{{$O_{\bigstar}=$} $\tfrac{1}{2}\sum_j \eta_j\, Z_j$ \\[4pt]
$\eta_j\!=\!\begin{cases}
+1 & j\!\leq\!\lfloor\frac{L}{4}\rfloor,\;
      j\!>\!\lfloor\frac{3L}{4}\rfloor \\[2pt]
-1 & \text{otherwise}
\end{cases}$}
& $\sim \frac{1}{12}L^2$ \\
\hline
\end{tabular}
\caption{Best operator and variance scaling of the Jastrow-Gutzwiller wave function for 
different values of $\alpha$.}
\end{table}

\section{Converging lower bounds to the Quantum Fisher Information}\label{sec:bounds}

We now introduce the two families of lower bounds we use to estimate the QFI for mixed states. In the next section we show how they can be efficiently evaluated via Monte Carlo sampling.

The first family, introduced in Ref.~\cite{Rath2021}, consists of polynomial lower bounds denoted with $F_n$. The key idea is to approximate the spectral function $f(\lambda_i, \lambda_j) = (\lambda_i - \lambda_j)^2/(\lambda_i + \lambda_j)$ appearing in the QFI (Eq.~\ref{eq:qfi}) by a polynomial in $\lambda_i$ and $\lambda_j$. This converts the double sum over eigenstates into combinations of traces involving mixed moments of $\rho$ and $O$
\begin{equation}\label{eq:bounds1}
    F_n=\sum_{k=0}^n 2^{k+1} \binom{n+1}{k+1}(-1)^k  T_k
\end{equation}
with 
\begin{equation}\label{eq:Tk}
T_k=\frac{1}{2^k} \sum_{l=0}^{k+2} C_l^{(k)} \operatorname{Tr}\left(\rho^{k+2-l} O \rho^l O \right),
\end{equation}
and
\begin{equation}
    C_l^{(k)}=\binom{k}{l}-2\binom{k}{l-1}+\binom{k}{l-2}.
\end{equation}
The bounds $F_n$ form a monotonically increasing sequence that converges to the exact QFI as $n \to \infty$. Furthermore $T_k$, and so $F_n$, can be accessed experimentally via randomized measurements without full state tomography~\cite{VitalePRX2024}. However, obtaining tight estimates of the QFI through this hierarchy becomes increasingly demanding: the experimental effort required to determine $F_n$ grows rapidly with $n$, which limits the practical accessibility of high-order bounds~\cite{VitalePRX2024}.

The second family, introduced in Ref.~\cite{Zhang2025}, is based on Krylov subspace methods, a class of techniques widely used in numerical linear algebra, most notably in the Lanczos algorithm. The central idea is to avoid diagonalizing a large matrix exactly: instead, one repeatedly applies the matrix to an initial vector, generating a sequence of vectors whose span defines the Krylov subspace. An optimal approximation to the desired solution is then sought within this restricted subspace. In our setting, the matrix is replaced by the linear super-operator $\mathcal{L}_{\rho}(A) = A\rho + \rho A$, which acts on operators $A$, and the initial vector is the commutator $i[\rho, O]$, encoding the infinitesimal response of the state to the generator $O$. Repeated application of $\mathcal{L}_\rho$ to this initial operator produces a hierarchy of operators spanning a Krylov subspace in operator space. Optimizing over linear combinations within this subspace yields a sequence of lower bounds on the QFI that become progressively tighter with the dimension of the subspace.
In this framework, one defines the matrix and the vector
\begin{equation}
    A_n=\left[\begin{array}{cccc}
T_1 & T_2 & \cdots & T_n \\
T_2 & T_3 & \cdots & T_{n+1} \\
\vdots & \vdots & \ddots & \vdots \\
T_n & T_{n+1} & \cdots & T_{2 n-1}
\end{array}\right], \quad b_n=\left[\begin{array}{c}
T_0 \\
T_1 \\
\vdots \\
T_{n-1}
\end{array}\right]
\end{equation}
where $T_n$ have been defined in Eq.~\eqref{eq:Tk}.
The corresponding Krylov bound is given by
\begin{equation}\label{eq:bounds2}
    B_n = b_n^T A_n^{-1} b.
\end{equation}
Like $F_n$, the Krylov bounds $B_n$ form a monotonically increasing sequence converging to $F_Q$, typically much faster than $F_n$.

Clearly, the two families are built from the same ingredients: the moments $T_k$. Fixing the moment budget up to the first $2n-1$ moments $T_0, \ldots, T_{2n-1}$ gives access to $F_{2n-1}$ and $B_n$.
It has been shown~\cite{Zhang2025} that
\begin{equation}
    F_{2n-1} \leq B_n \leq F_Q,
\end{equation}
so the Krylov bound is always at least as tight as the polynomial bound computed from the same data. The gap between either bound and the true QFI closes exponentially in $n$, but with a faster rate for $B_n$.

The convergence of both families is controlled by the spectral structure of $\rho$. When the density matrix has low effective rank, i.e. its eigenvalue distribution is concentrated on a small number of values, a few moments $T_k$ already capture the dominant contributions to the QFI, and both bounds converge rapidly. 
In contrast, when $\rho$ is high-rank, the same finite set of moments must resolve contributions from a much larger number of independent transitions, leading to slower convergence and systematically looser bounds at any fixed order $n$.
The values of $B_n$ suffer from an additional numerical bottleneck. With increasing $n$, the determinant of the matrix $A_n$ of Eq.~\eqref{eq:bounds2} rapidly approaches zero, due to column vectors becoming nearly linearly dependent. Inverting $A_n$ then requires estimates of $T_k$ with precision smaller than the determinant itself. In practice, this means that, for most states and operators, estimating $B_n$ for $n > 2$ becomes unfeasible.
Both families perform best when $\rho$ has low effective rank or is highly mixed but, as we show in the next section (Sec.~\ref{sec:QFIwithdecoherence}), the structure of the operator $O$ relative to the noise channel can also condition the convergence at high rank by restricting the transitions that contribute to the QFI as mentioned above.

\section{Quantum Fisher information of noisy Jastrow-Gutzwiller wave functions}\label{sec:QFIwithdecoherence}

We now turn to the central question of this work: evaluating the QFI of JG wave functions in the presence of decoherence by exploiting the explicit knowledge of the JG wave function amplitudes in the occupation number basis.
The starting point is the pure-state density matrix
\begin{equation}
    \rho_0 = \ket{\Psi_\alpha}\bra{\Psi_\alpha} = \sum_{n n'} c_n c_{n'}\ket{n}\bra{n'}
\end{equation}
where $c_n = \psi_\alpha(\{n\})$ and $\ket{n} = \ket{n_1 n_2 \dots n_L}$. 

We consider the action of a quantum channel $\mathcal{E}$ on this state $\rho = \mathcal{E}[\rho_0]$.
The output noisy density matrix retains a structured form in the Fock basis,
\begin{equation} 
\rho = \sum_{n,n'} \tilde{c}_{n,n'}\, \ket{n}\bra{n'}, \end{equation}
that can be connected to the original amplitudes $c_n$ through the channel parameters.
The key observation is that, despite the state being mixed and the resulting expressions being nontrivial, the traces entering the lower bounds in Eqs~\eqref{eq:bounds1} and~\eqref{eq:bounds2} can be expanded as sums over products of the original squared amplitudes $c_n^2$, multiplied by weights that depend on the specifics of the noise channel $\mathcal{E}$ applied and on the operator $O$ employed.

Since the distribution $\{ c_n^2 \}$ defines a normalized probability distribution over the Fock-space configurations at half filling, these sums are expectation values of an observable over a classical probability distribution, and can therefore be estimated efficiently using Markov chain Monte Carlo sampling.
The computational cost of this approach is analyzed in detail in App.~\ref{app:montecarlo}.

In the following, we study three noise channels, chosen to span qualitatively different relationships between the channel, the operator $O$ and the resulting density-matrix rank. We begin with local dephasing (Sec.~\ref{sec:local_dephasing}), which preserves the diagonal of the density matrix and only suppresses off-diagonal coherences. We then consider amplitude damping (Sec.~\ref{sec:local_amp}), which modifies both diagonal and off-diagonal elements. Finally, we treat the global depolarizing channel (Sec.~\ref{sec:global_dep}), which uniformly mixes the state toward the maximally mixed state. 
In each case, we derive the explicit form of the decohered density matrix, show how the relevant traces reduce to Monte Carlo amenable expressions, and present numerical results for the polynomial and Krylov bounds to the QFI across different values of $\alpha$ and noise strength $p$. 
For more details about the analytics we refer the reader to App.~\ref{app:dephasing} and ~\ref{app:damping}.

\subsection{Local dephasing}\label{sec:local_dephasing}
The local dephasing channel can be described as a composition of Pauli  $Z$ quantum channels acting independently on each site. Its action on the density matrix can be written as
\begin{equation}\label{eq:Zchannel}
    \rho_0\to \rho =\prod_j \mathcal{E}^Z_j[\rho_0].
\end{equation}
Here $\mathcal{E}^Z_j(\rho_0)=(1-p)\rho_0+p Z_{j}\rho_0 Z_{j} $ and $p$ is the decoherence strength. Since we know the explicit form of the state in the occupation-number basis, the action of the channel can be computed straightforwardly. Each local $Z_j$ acts diagonally multiplying the state by a phase factor depending on the local occupation number. As a consequence, the diagonal elements of the density matrix remain unchanged, while the off-diagonal terms acquire relative phase factors. Applying the channel on all sites leads to the decohered density matrix (see App.~\ref{app:loc_deph_analytical})
\begin{equation}\label{eq:state_local_dephasing}
    \rho_p = \sum_{n n'} c_n c_{n'} (1-2p)^{N''(n, n')} \ket{n}\bra{n'}
\end{equation}
with $N''=\sum_i\left[ (n_i + n_i')\operatorname{mod} 2\right]$. In this form, it is clear that decoherence leaves the populations $c_n^2$ unaffected but exponentially suppresses coherences between configurations that differ on many sites. 

Thanks to this explicit representation, the traces entering the bounds~\eqref{eq:bounds1} and~\eqref{eq:bounds2} can be written as (for $r,s>0$)
\begin{equation}
\label{eq:Mc_traces}
\begin{aligned}
&\operatorname{Tr}\left(\rho_p^r O \rho_p^s O\right)
= \sum_{\substack{n_1\dots n_{r+1} \\\ m_1\dots m_{s+1}}} c_{n_1}^2 c_{n_2}^2  \dots c_{n_r}^2 
c_{m_1}^2 c_{m_2}^2  \dots c_{m_s}^2 \\ &
f_{n_1 n_2}  \dots f_{m_s m_{s+1}} \frac{c_{n_{r+1}} c_{m_{s+1}}}{c_{n_1} c_{m_1}}\langle n_{r+1}|O|m_1\rangle
\langle m_{s+1}|O|n_1\rangle 
\end{aligned}
\end{equation}
with $f_{nn'}=(1-2p)^{N''}$.
The factors $c_n^2$ are the probability distribution of the original JG wave function (Eq.~\eqref{eq:wf}). This allows us to evaluate the traces using Monte Carlo sampling since the configurations can be generated via a Markov-chain algorithm sampling the distribution $c_n^2$, while the remaining factors in Eq.~\eqref{eq:Mc_traces} are evaluated as weights along the sampled trajectories. 

Let us now break down the behavior of the QFI of the JG as a function after local dephasing as a function of the parameter $\alpha$, following the operator choices summarized in Table~\ref{table:var}.
\begin{figure*}[ht]
    \centering
    \includegraphics[scale=.42]{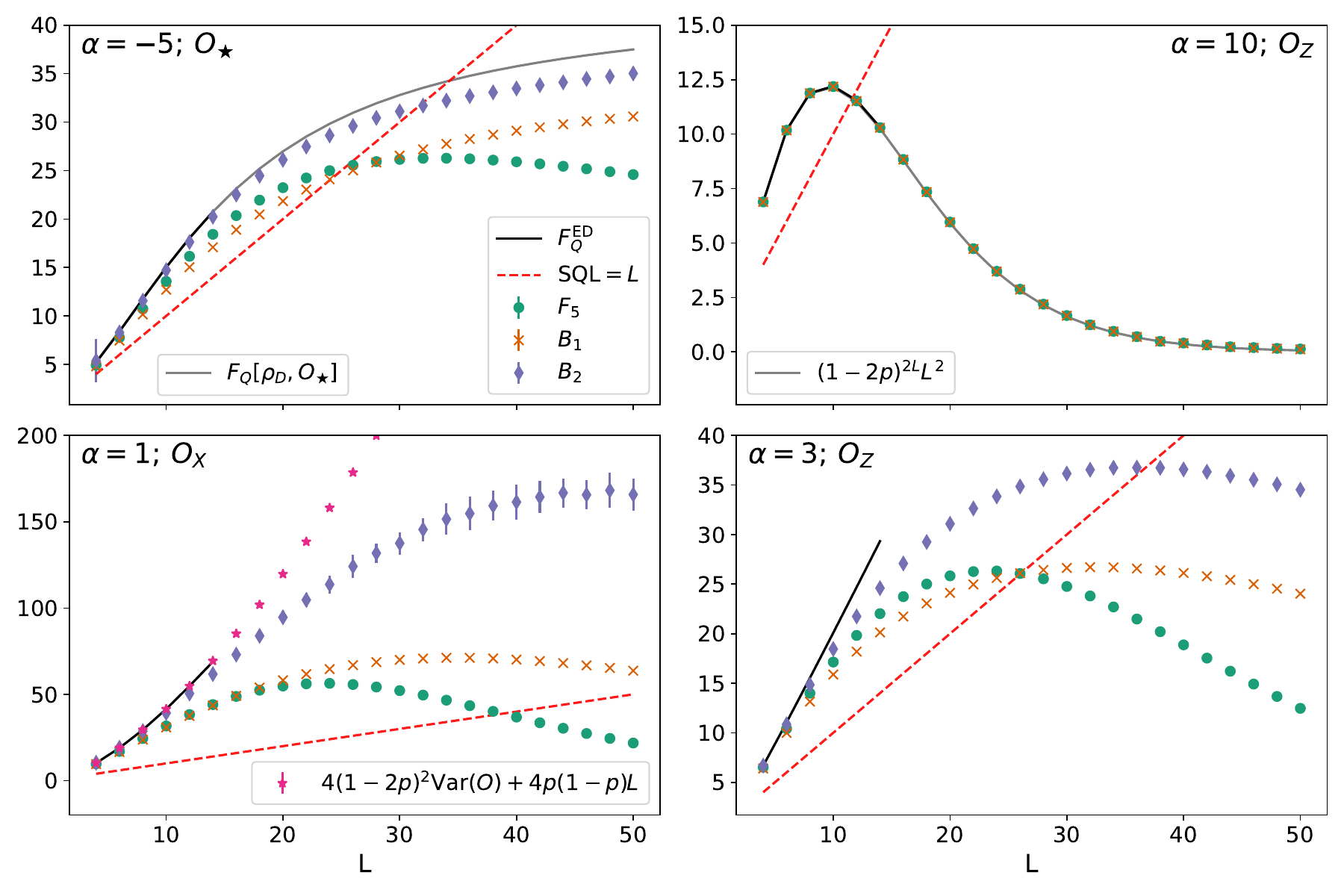}
    \caption{Numerical results on the bounds to the QFI for the dephased JG wave function for different values of $\alpha$, corresponding to different regimes of the wave function, with local dephasing probability $p = 0.05$. The operators considered for the different values of $\alpha$ are the ones reported in Table~\ref{table:var}. The dashed line represents the SQL, i.e., the threshold that signals the presence of entanglement in the state. The solid line corresponds to the exact QFI, obtained via exact diagonalization for small system sizes. For $\alpha=1$ we also include the results for the values of the real QFI, by numerically computing the variance of the operator and get the QFI using Eq.~\ref{eq:qfi_exact} (the results for $L > 30$ are cut for visualization convenience), while the gray lines for $\alpha = -5, 10$ are the exact QFI for the limit cases of the GHZ state and of the ``Dicke-like" state~\ref{eq:cdicke}, that we derive in App.~\ref{app:dephasing}. Where we can compare the bounds to the exact QFI, we see that when the density matrix is low-rank (for $\alpha =-5, 10$) and the bounds approximate well the QFI, while the opposite is true for the critical states. Still, for the system sizes and values of $\alpha$ considered, the Krylov bounds $B_1$ and $B_2$ work as entanglement witnesses.}\label{fig:dephasing}
\end{figure*}
\paragraph{\bf $(i)$ $\alpha > 4$, operator $O_Z$.} 
For $\alpha >4$, the optimal operator is the  staggered magnetization $O_Z$. Since  $O_Z$ is diagonal in the occupation-number basis, 
\begin{equation}
    \langle n|O_Z|m\rangle = (O_Z)_{nn}\, \delta_{nm},
\end{equation}
the matrix elements in Eq.~\eqref{eq:Mc_traces} enforce the 
constraints $m_1 = n_{r+1}$ and $m_{s+1} = n_1$, collapsing the sums over $n_{r+1}$ and $m_{s+1}$. The Monte Carlo integration is particularly efficient in this case, since the operator does not induce transitions between configurations.

In the strict GHZ limit ($\alpha \gg 1$), the dephased QFI can be computed analytically. Local dephasing suppresses the coherence between  $\ket{101\dots10}$ and $\ket{010\dots01}$, and since $O_Z$ is diagonal, the QFI receives contributions only from these terms, yielding (see App.~\ref{app:dephasing})
\begin{equation}\label{eq:qfidephasing}
    F_Q[\rho_p, O_Z] = L^2 (1-2p)^{2L}.
\end{equation}
The QFI is therefore exponentially suppressed in $L$ at any finite $p$.

\paragraph{\bf $(ii)$ $\alpha \in (2,4]$, operator $O_Z$.} In this region, the optimal operator remains $O_Z$ but the bounds no longer have a simple analytical form.
An alternative way to extract at least the scaling of the QFI with system size $L$ is by constructing a bound based on error propagation~\cite{toth2014quantum,Sara}, as shown in App.~\ref{app:sarabound}. The resulting bound shows that, in the presence of noise, the QFI scales at least linearly with system size, and thus returns to the SQL. This result still represents an improvement over the GHZ-like regime $\alpha\gg 1$, where the QFI is exponentially suppressed with $L$ at any finite noise strength, see Eq.~\eqref{eq:qfidephasing}.

\paragraph{\bf $(iii)$ $\alpha \in [0,2]$, operator $O_X$.} Now the optimal generator is $O_X = \tfrac{1}{2}\sum_j (-1)^j X_j$, which connects different occupation-number configurations. The trace
$\mathrm{Tr}(\rho^r O\rho^s O)$ is nonzero only when either $r=0$ or $s=0$. For $r,s>0$, Eq.~\eqref{eq:Mc_traces} vanishes because $O_X$ is off-diagonal in the occupation-number basis and therefore has no matrix elements that can contribute to the trace over half-filling configurations. We refer to App.~\ref{app:dephasing} for further details. 

Furthermore, when $O=O_{X}$, one can identify a discrete $\mathbb{Z}_2$ subgroup of the $U(1)$ symmetry generated by the parity operator $S=\prod_j Z_j$. This symmetry anti-commutes with $O_{X}$ ($S O_{X} S^{-1}=-O_{X}$), while commuting with the local dephasing channel $\mathcal{E}_j^Z$, so the $\mathbb{Z}_2$ symmetry is preserved by the noise. As shown in Ref.~\cite{Sara}, when noise channels are strongly $\mathbb{Z}_2$-symmetric, i.e. they preserve the symmetry generated by $S$, and the symmetry generator anti-commutes with $O$, the asymptotic scaling of the QFI remains unchanged. The effect of the noise is therefore limited to a renormalization of the overall pre-factor, while the scaling with system size is preserved. The QFI for this specific mixed state is given by 
\begin{equation}\label{eq:qfi_exact}
    F_{Q}[\rho_p,O]=4(1-2p)^2\mathrm{Var}_\rho(O)+4p(1-p)L,
\end{equation}
where the variance is evaluated in the pristine Jastrow-Gutzwiller wavefunction, and can be easily obtained using the Monte Carlo calculation described before, by setting $r=2$ and $s=0$ in Eq.~\eqref{eq:Mc_traces}, for considerable sizes of the system.

\paragraph{\bf $(iv)$ $\alpha < 0$, operator $O_\bigstar$.} Like 
$O_Z$, $O_\bigstar$ is diagonal in the occupation-number basis and the same estimator simplification applies. In the strongly clustered limit $\alpha \ll -1$, the state is well approximated by the Dicke-like superposition of Eq.~\eqref{eq:cdicke}, and the resulting density matrix is again low-rank. In contrast to the GHZ state, the growing number of coherences contributing to the QFI as the system size increases can compensate for their exponential suppression by dephasing. In App.~\ref{qfidicke} we derive an analytical formula for the QFI of the state and show that in the thermodynamic limit the QFI tends to a $p$-dependent constant (independent of $L$).


\begin{figure*}[ht]
    \centering
    \includegraphics[scale=.42]{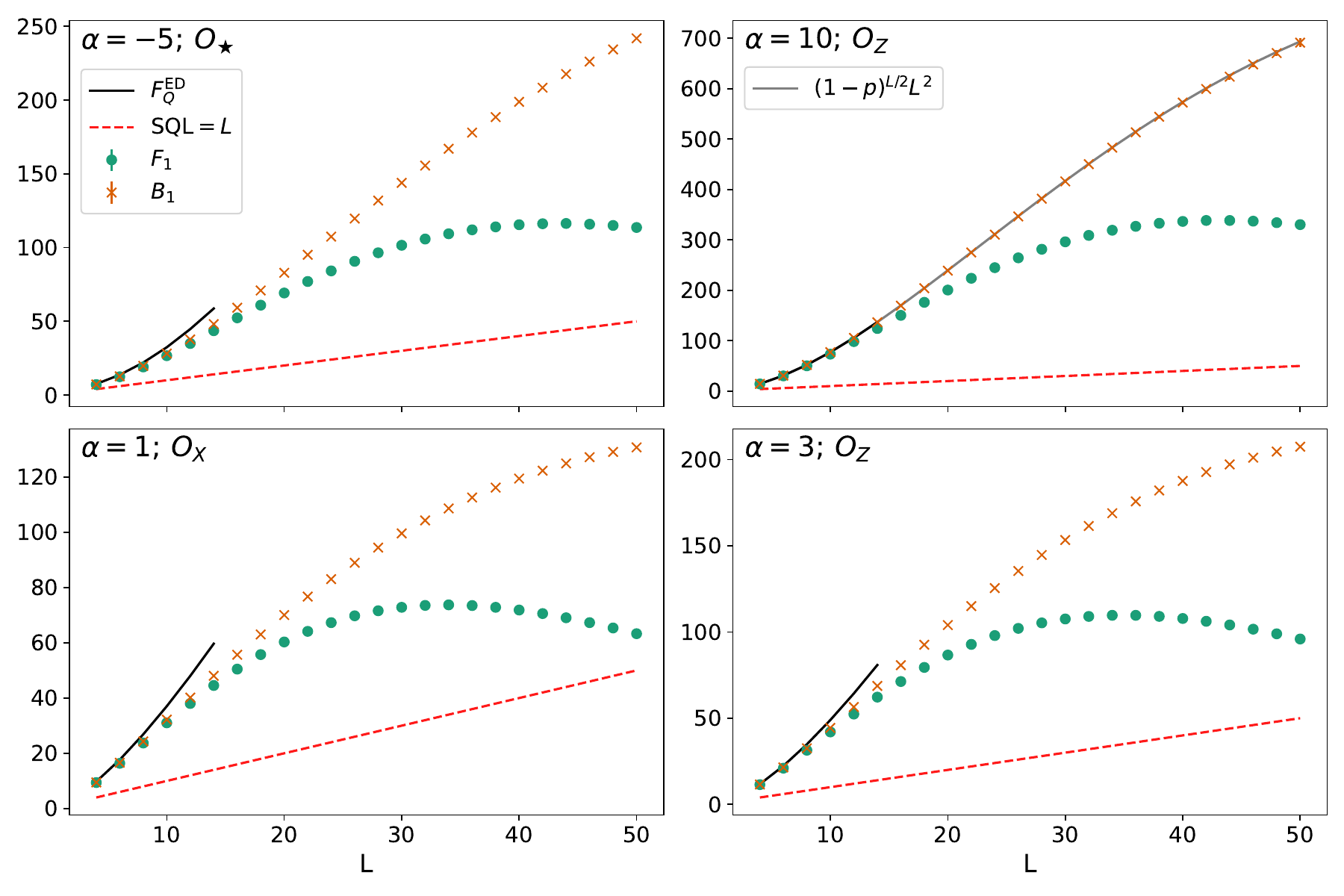}
    \caption{Numerical results on the bounds to the QFI for the damped JG wave function for different values of $\alpha$, corresponding to different regimes of the wave function with decoherence strength $p = 0.05$. The operator considered for the different values of $\alpha$ are the ones reported in Table~\ref{table:var}. The dashed line represents the SQL, i.e., the threshold that signals the presence of entanglement in the state. The solid line corresponds to the exact QFI, obtained via exact diagonalization for small system sizes. For $\alpha = 10$, $B_1$ immediately saturates to the QFI (solid line) because the operator $O_Z$ is diagonal in the occupation basis and effectively reduces it to a low-rank state in the computation of the QFI.}\label{fig:damping}
\end{figure*}
In Fig.~\ref{fig:dephasing}, we show the Monte Carlo results for the different regimes discussed above, after applying the local dephasing channel in Eq.~\eqref{eq:Zchannel} with $p=0.05$, and for several values of $\alpha$. We choose $p=0.05$ as a representative non-perturbative noise strength: for the system sizes considered here, $pL$ is of order unity, so that the noise is strong enough to substantially modify the density-matrix spectrum, while still preserving a nontrivial metrological content. For small system sizes, we report the true QFI obtained via exact diagonalization, that is well captured by the lower bounds, which become less tight as the system size increases. For $\alpha = 1$, we include the exact QFI for all system sizes. In this case, we use the operator $O_{X}$, for which Eq.~\eqref{eq:qfi_exact} applies, so that the QFI can be estimated by computing the pure-state variance via Monte Carlo sampling. Whenever possible, we also report the analytical predictions for the QFI. The red dashed line marks the SQL, allowing us to identify the regimes in which the states retain a metrological advantage and multi-partite entanglement.

Let us comment on the agreement between our bounds and the exact value of the QFI. Since local dephasing acts only with operators that are diagonal in the occupation number basis, a large spread in configuration space of the pure state corresponds directly to a high rank density matrix for the mixed state. As described above, this translates to bounds on the QFI that are tighter on states where the rank is lower: for $\alpha = 10$, where the state is essentially GHZ-like, the bounds saturate the QFI. Also for $\alpha=-5$, $B_2$ is pretty close to the true value of the QFI (black solid line) for the system sizes we have considered. In contrast, for critical states, the bounds provide a poorer approximation to the actual QFI. Even in the latter case, the Krylov bounds can work as a cheap entanglement witness: in the case $\alpha =1$, $B_1$ is always above $L$ across all system sizes considered here, which indicates that the state is entangled and more useful, compared to classical states, for quantum metrology. 
We also highlight the computational advantage of the Krylov bounds $B_n$ as system size grows. From a numerical standpoint, $F_5$ and $B_2$ are related to the same number of traces of the form in Eq.~\eqref{eq:Mc_traces}.
However, we find that for larger systems, even $B_1$ already provides a substantially tighter bound on the true QFI than $F_5$. 
Remarkably, computing $B_1 = T_0^2 / T_1$ only requires traces up to $\operatorname{Tr}(\rho^3 O^2)$, while $F_5$ (or $B_2$) requires going up to $\operatorname{Tr}(\rho^7 O^2)$. This makes $B_1$ a much more easily accessible quantity, including when the channel is not diagonal in the occupation-number basis, as in the next section.

\begin{figure*}[ht]
    \centering
    \includegraphics[scale=.42]{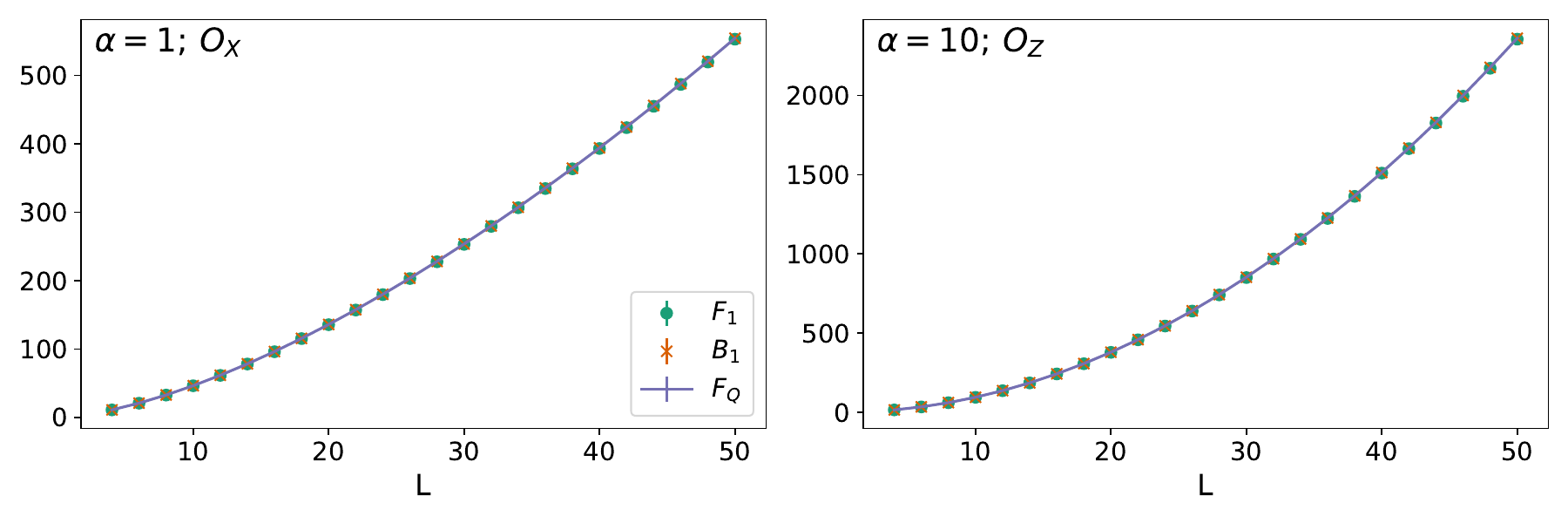}
    \caption{Numerical results for the bounds on the QFI of the depolarized JG wave function for $\alpha=1,10$. For each value of $\alpha$, we consider the operator reported in Table~\ref{table:var}, using the corresponding symbol. Here $p=0.05$ denotes the strength of the depolarizing channel.}\label{fig:depolarizing}
\end{figure*}
\subsection{Local amplitude damping}\label{sec:local_amp}

In contrast to local dephasing, amplitude damping describes an incoherent process in which an excitation can be lost to the environment. This means that an empty site $|0\rangle_j$ is left unchanged, while the state
$|1\rangle_j$ is either preserved with reduced amplitude or converted into
$|0\rangle_j$. Thus the channel not only suppresses quantum coherences, but also changes the occupation configuration itself. The full channel is obtained by applying the following Kraus operators on every site
\begin{equation} K_j^{(0)} = |0\rangle\langle 0|_j + \sqrt{1-p}\,|1\rangle\langle 1|_j, \qquad K_j^{(1)} = \sqrt{p}\,|0\rangle\langle 1|_j. 
\end{equation}
The parameter $p$ controls the relaxation strength:
for $p=0$ the state is unchanged, while for $p=1$ all occupied sites are mapped to empty sites. This implies that if the state at $p=0$ has a fixed number of particles, the amplitude damping produces a mixed state with support in several particle-number sectors. As shown in App.~\ref{app:damping}, a Monte Carlo estimator can also be constructed in this case to evaluate the building blocks~\eqref{eq:Tk} entering the lower bounds to the QFI. However, the structure of the estimator is more involved than for local dephasing, for the reasons above. 
Consequently, the corresponding Markov chain must keep track not only of the classical probabilities associated with the damping process, but also of the wave function amplitudes of the configurations connected by the channel. We give the explicit construction in App.~\ref{app:damping}. Due to this additional complexity, we restrict ourselves to the evaluation of $F_0$, $F_1$, and $B_1$. The results are reported in Fig.~\ref{fig:damping}. 
As in the local dephasing case, for $\alpha = 10$ the state is well approximated by a GHZ state, and the bounds saturate the QFI. However, in contrast to the dephasing channel, here the noisy state is not low-rank, since the noise connects states in all sectors with a smaller particle number. In this case, the bounds converge immediately to the QFI because of the specific operator considered. Indeed, as we noticed for local dephasing, since the operator $O_Z$ acts diagonally in the occupation number basis, the only contribution to the QFI comes from the coherence between the configurations $\ket{101 \ldots 10}$ and $\ket{010 \ldots 01}$. The amplitude-damping channel suppresses this coherence by a factor $(1-p)^{L/2}$ because only the $L/2$ occupied sites contribute a factor $\sqrt{1-p}$ in each GHZ component. Therefore, the final result for the QFI is (see App.~\ref{app:damping} for more details)
\begin{equation}\label{eq:qfidamping}
    F_{Q}[\rho_p,O_Z]=L^2(1-p)^{L/2}.
\end{equation}
The Heisenberg scaling $L^2$ is exponentially suppressed in $L$ for any finite damping strength $p>0$, reflecting the fragility of GHZ coherence under amplitude damping. Remarkably, the bounds converge to the QFI since the state remains effectively low-rank, given the operator considered. Away from the GHZ regime, our bounds reveal a substantial metrological advantage over the SQL. However, already at small system sizes, where the exact QFI is accessible via exact diagonalization, we observe a deviation between the bounds and the true QFI.

\subsection{Global depolarizing}\label{sec:global_dep}

Let us now consider a global depolarizing channel, which acts on the pure state $\rho$ as
\begin{equation}
    \rho_p = (1-p) \rho + \frac{p}{2^L}\mathbb{I}.
\end{equation}
In this case, the derivation of the QFI is straightforward and we have~\cite{Rath2021}
\begin{equation}\label{eq:qfidepol}
    F_Q[\rho_p, O] = \operatorname{Var}_\rho(O) \frac{(1 - p)^2}{1 - p + p / 2^{L-1}}
\end{equation}
where the variance of the operator is computed on the pure state $\rho$. Since the QFI is given by the variance multiplied by an overall factor, the QFI follows the same scaling as the variance reported in Table~\ref{table:var}, for each value of $\alpha$ and its corresponding optimal operator. 
Although the effect of this channel on the exact QFI turns out to be very simple, it is still interesting to analyze the performance of the QFI lower bounds. Indeed, the depolarizing channel drives the state toward the maximally mixed state, making the spectrum of $\rho$ flat. For this reason, the bounds quickly converge to the true value of the QFI in Eq.~\eqref{eq:qfidepol} for any initial state. 
The bounds for this channel can be obtained starting from the variance of pure state, noting that
\begin{equation}
    \rho_p^n  =\left( \left(q + t\right)^n - t^n\right) \rho + t^n \mathbb{I}
\end{equation}
where we defined $ q = 1 -p$ and $t = \tfrac{p}{2^L}$, and defining $u^{(n)} = \left(q + t\right)^n - t^n$ we have
\begin{equation}
    \begin{aligned}
        \operatorname{Tr}(\rho^r O \rho^s O) = &\left(u^{(r)} t^s + u^{(s)} t^r\right) \braket{O^2}_\rho + \\
        &+ u^{(s)} u^{(t)} \braket{O}_\rho^2 + t^r t^s \operatorname{Tr}(O^2).
    \end{aligned}
\end{equation}
We show the QFI and its bounds in Fig.~\ref{fig:depolarizing}. The (almost completely) flat spectrum of the density matrix makes it effectively low-rank with regard to the QFI (where all the terms with $\lambda_i =\lambda_j$ and $\braket{i|j} = 0$ with $i \neq j$ vanish), which results in the bounds $F_n$ and $B_n$ already converging to the QFI for $n = 1$. We show it only for two arbitrary values of $\alpha$, but the result does not depend on the specific initial pure state.

\section{Discussion and outlook}

\begin{table*}[t]
\centering
\setlength{\tabcolsep}{6pt}
\begin{tabular}{|l|c|c|c|c|}
\hline
\textbf{State / operator} & \textbf{Pure state} & \textbf{Local dephasing} & \textbf{Amplitude damping} & \textbf{Depolarizing} \\
\hline
GHZ-like ($\alpha\gg 1$, $O_Z$)
& $F_Q \sim L^2$
& $F_Q \simeq L^2(1-2p)^{2L}$
& $F_Q \simeq L^2(1-p)^{L/2}$
& $F_Q \sim (1-p) L^2$
\\
\hline
Dicke-like ($\alpha\ll -1$, $O^\star$)
& $F_Q \sim L^2 $
& $F_Q \sim 1/p^2$
& - - -
& $F_Q \sim (1-p) L^2$
\\
\hline
Critical ($0\leq\alpha\le 2$, $O_X$)
& $F_Q \sim L^{2-\alpha/2}$
& $F_Q \sim 4(1-2p)^2 L^{2-\alpha/2}$
& - - -
& $F_Q \sim (1-p) L^{2-\alpha/2}$
\\
\hline
Critical ($2<\alpha\lesssim 4$, $O_Z$)
& $F_Q \sim L^{2-2/\alpha}$
& $F_Q \gtrsim \mathcal{O}(L)$
&  - - -
& $F_Q \sim (1-p) L^{2-2/\alpha}$
\\
\hline
\end{tabular}
\caption{Summary of QFI behavior across Jastrow--Gutzwiller regimes and noise channels, where $p$ denotes the noise strength. Whenever possible, we report the explicit analytical results.}
\label{tab:summary_noise_regimes}
\end{table*}

In this work, we have investigated the metrological properties of JG wave functions, which interpolate between GHZ-like and critical regimes as a function of the effective interaction between occupied sites (parametrized by $\alpha$). We first characterized the behavior of the QFI for pure states, identifying, in each regime, the operators that yield the largest scaling with system size. For large positive values of $\alpha$, the state approaches an antiferromagnetic GHZ state and displays Heisenberg scaling with the staggered magnetization. In the critical regime, the scaling of the QFI is controlled by the algebraic decay of correlation functions, and the optimal operator is $\alpha$ dependent. For negative $\alpha$, where the wave function favors clustered configurations, a more structured magnetization operator again leads to Heisenberg scaling.  

To have access to the QFI under the action of different types of noise channels, we have taken advantage of the explicit expression of JG wavefunction. We have shown that lower bounds of the QFI (Ref~\cite{Rath2021,VitalePRX2024,Zhang2024,Zhang2025} can be computed as averages over the classical probability distribution defined by the wave function amplitudes, which allows us to access QFI estimates for system sizes well beyond exact diagonalization. The performance of the bounds depends strongly on the structure of the noisy density matrix. 
For local dephasing, the bounds are very accurate in the GHZ-like regime, where the effective rank of the state remains small, and they become less tight when the wave function has support over many configurations and the noisy density matrix has a much larger effective rank. Nevertheless, even when the bounds do not converge to the exact QFI, they can provide useful information, e.g. estimating a value above the SQL and witnessing the metrological utility of the state even in presence of noise. 
In the case of local amplitude damping, the Monte Carlo estimators are more involved because the noise channel modifies the particle-number sector. Nevertheless, we showed that the bounds are very tight even though the matrix is not low-rank, thanks to the structure of the operator $O$. 
Finally, for the global depolarizing channel, the exact QFI reduces to the pure-state variance multiplied by a noise-dependent factor. This channel, therefore, provides a good benchmark where the bounds approach the exact result, consistently with the fact that the state is driven toward a highly mixed form. Table~\ref{tab:summary_noise_regimes} summarizes the scaling of the QFI in the different regimes of the JG wave function and for the noise channels we have considered. We emphasize that both the GHZ-like phase $\alpha>4$ and the Dicke-like phase $\alpha<0$ are particularly fragile to local noise. Indeed, in the regimes where we have analytical control, we show that the QFI either decreases exponentially with system size or saturates to an $L$-independent value. By contrast, in the critical regime, even in the worst case, the QFI is reduced only to the SQL.

Our work opens several directions for future exploration. First, the strategy developed here is quite general: whenever an explicit representation of the many-body wave function is available, information-theoretic quantities can, in principle, be evaluated by mapping them onto properly defined Markov chains. This opens the possibility of applying similar methods to other analytically known wave functions, such as Laughlin or fractional quantum Hall states, also in higher dimensions. For example, one could study the behavior of quantities such as the quantum coherent information or the purity, which have already been investigated in related settings~\cite{Wang2025}. Another natural question concerns the preparation of the JG states studied here. While these wave functions provide a useful analytically controlled family of metrological resources, it remains important to identify realistic protocols to prepare them on near-term quantum devices~\cite{xu2025}. In this direction, it would be particularly interesting to understand whether the regimes that retain a metrological advantage under noise can be prepared more efficiently than the GHZ-like regime, whose QFI is extremely fragile under local decoherence. An interesting follow-up would be to evaluate the Krylov bounds, $B_n$, using randomized measurements: they are defined in terms of the moments $\mathrm{Tr}(\rho^r O \rho^s O)$, which can be accessed through shadow-tomography protocols. This was done for the bounds $F_n$ in~\cite{Rath2021}, but since $B_n$ provide a tighter estimation of the QFI, it would be interesting to have experimental access to at least the first moments. Another question, closely related to quantum metrology, concerns the identification of the best observable to measure in order to practically extract $\delta\theta$. In Ref.~\cite{Sara}, a symmetry-based algorithm was used to identify optimal measurement strategies, mainly relying on discrete symmetries. It would be interesting to understand whether continuous symmetries, such as the $U(1)$ symmetry present here, can play an important role in this analysis as well.

\section*{Acknowledgements}
VV is grateful to Aniket Rath for useful discussions and previous collaborations. SM acknowledges inspiring discussions with Jason Alicea, Yinan Chen, Kabir Khanna, Pablo Sala, and Romain Vasseur. We also thank Silvia Pappalardi and Xhek Turkeshi for their comments on the draft.

\bibliography{references}

\clearpage
\onecolumngrid
\appendix

\section{Correlations and variances of the Jastrow-Gutzwiller wave function in the critical regime}
\label{app:luttinger}

We want to identify the operator with the best scaling of the variance with system size. This is directly related to metrology, since for pure states the variance is proportional to the QFI. For the JG wave function of Eq.~\ref{eq:wf}, the optimal operator can be easily obtained in the regimes $\alpha<0$ and $\alpha>4$, where the structure of the wave function is particularly simple. In the critical phase, instead, the choice is less obvious.

To address this regime, we use the fact that the critical Jastrow--Gutzwiller wave function is described by a Luttinger liquid with Luttinger parameter $K=1/\alpha$~\cite{Turkeshi2020}. In a Luttinger liquid, the asymptotic decay of correlation functions can be obtained using standard bosonization techniques, for instance, starting from the XXZ spin chain with positive interaction~\cite{Gogolin2004}. Using the dictionary between the Pauli spin operators and the bosonic fields, the resulting correlation functions are given by~\cite{Gogolin2004}
\begin{equation}\label{eq:corr}
\begin{aligned}
    \langle Z_j Z_0 \rangle &\sim 4 m^2 + \frac{1}{2 \pi \alpha} \frac{1}{j^2} + \frac{2}{\pi^2}(-1)^j \frac{1}{j^{2/\alpha}} \\
    \langle X_j X_0 \rangle = \langle Y_j Y_0 \rangle &\sim |\mathcal{C}|^2 (-1)^j\frac{1}{j^\frac{1}{2K}} + |\mathcal{A}|^2  \frac{1}{j^{2K + \frac{1}{2K}}}
\end{aligned}
\end{equation}
where the constants $\mathcal{C}$ and $\mathcal{A}$ are not universal. Therefore, the variances of the staggered $X$ and $Z$ operators are
\begin{equation}
\label{eq:vars}
\begin{aligned}
    \operatorname{Var}\bigg(\sum_j (-1)^j Z_j\bigg) &= L +  L \sum_{j=1}^{L-1}(-1)^j\langle Z_jZ_0\rangle \sim L + L  \int_\varepsilon^L \frac{dr}{r^{\frac{2}{\alpha}}} \sim
    \begin{cases}
        L & \text{if}\ \alpha < 2 \\
        L^{2-2/\alpha} & \text{if}\ \alpha > 2
    \end{cases}\\
    \operatorname{Var}\bigg(\sum_j (-1)^j X_j\bigg) &= L +  L \sum_{j=1}^{L-1}(-1)^j\langle X_jX_0\rangle \sim L + L \int_\varepsilon^L \frac{dr}{r^{\frac{\alpha}{2}}} \sim 
    \begin{cases}
        L^{2-\alpha/2} & \text{if}\ \alpha < 2 \\
        L & \text{if}\ \alpha > 2
    \end{cases}
\end{aligned}
\end{equation} 
Using this dictionary, we derive the correlators and the variance of the Pauli spin operators also in the JG wave functions.
\begin{figure}[h]
    \centering
    \includegraphics[scale=.44]{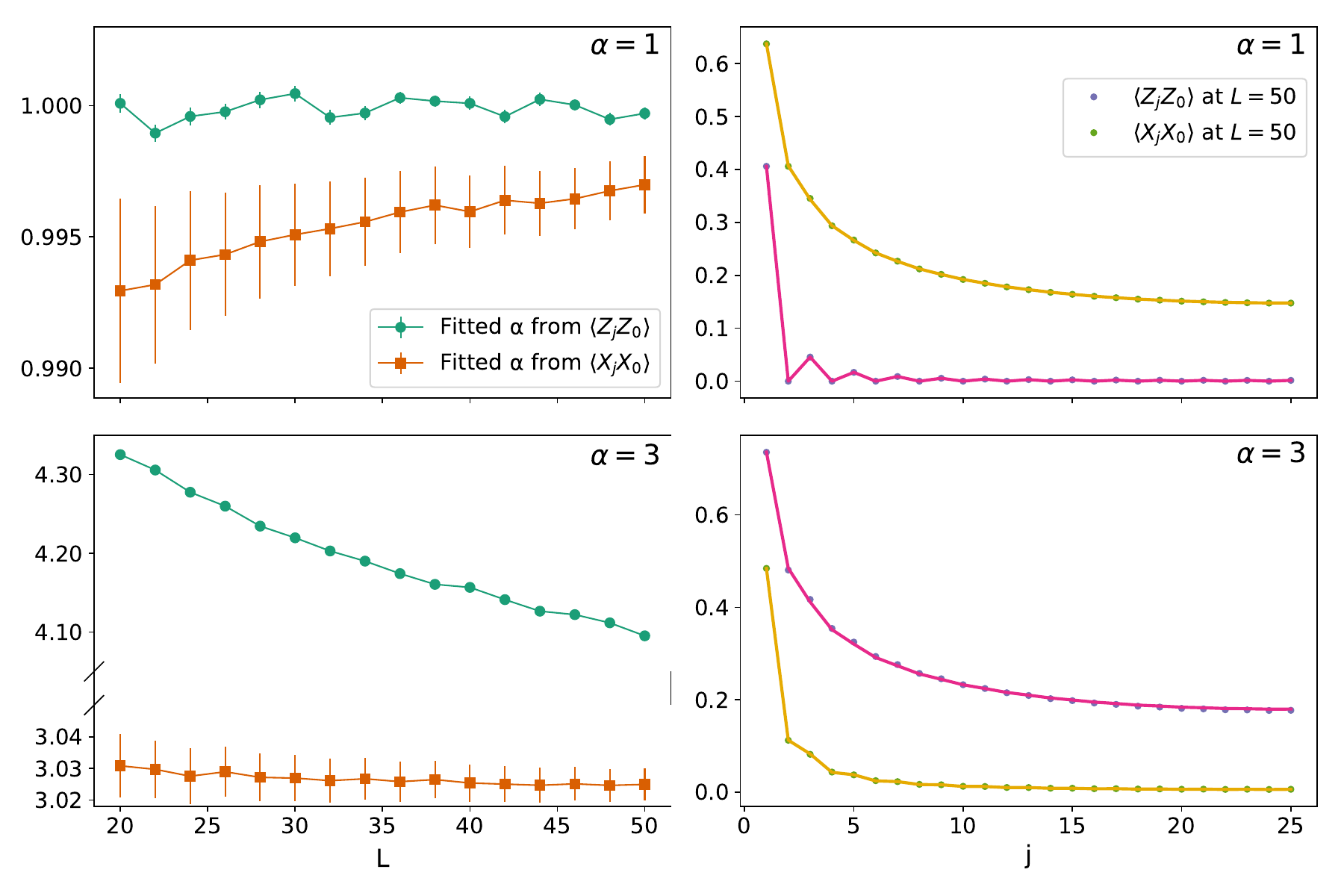}
    \caption{Correlation function of the JG wave functions in the critical regime. For $\alpha = 1, 3$ we computed the correlation functions of $Z$ and $X$ for every $L$ and fitted the results with expected behavior from Eq.~\ref{eq:corr} to check the agreement between the value of $\alpha$ use to construct the wave function and the one coming from the fit. In the left column the value of $\alpha$ coming from the fit is plotted against the system size, while on the right we report the values of the correlations at $L=50$ with the corresponding fits.}
    \label{fig:corrs}
\end{figure}

To verify these predictions, we compute the correlation functions for $\alpha=1$ and $\alpha=3$ up to system size $L=50$. The numerical results are obtained by Monte Carlo sampling of the probability distribution $|\psi_\alpha(\{n\})|^2$, as described in the next section. We then fit the data using the asymptotic forms in Eq.~\eqref{eq:corr}. In the fits we set $m=0$, corresponding to half filling, and treat the coefficients and the effective value of $\alpha$ as fitting parameters.

The fitted values of $\alpha$ obtained from the $X$ and $Z$ correlation functions are shown in Fig.~\ref{fig:corrs} for each system size. The $X$ correlator is in good agreement with the expected behavior for both $\alpha=1$ and $\alpha=3$. When $\alpha=3$, the $Z$ correlator shows stronger finite-size effects: the fitted value of $\alpha$ decreases with increasing $L$, but remains squite larger than the expected asymptotic value over the range of sizes accessible numerically.
 
\section{Sampling of the Jastrow-Gutzwiller wave function distribution}\label{app:montecarlo}

To obtain, via Monte Carlo integration, the bounds on the QFI described in Sec.~\ref{sec:QFIwithdecoherence}, we need to be able to sample from the probability distribution given by the JG wave function. To do so, we employ a simple Markov chain to get $M$ samples from the $c_n^2$ ($\psi^2_a(\{n\}$) of Eq.~\ref{eq:wf} for every value of $\alpha$ and $L$ we are interested in. Then to compute the traces $\operatorname{Tr}(\rho^r O \rho^s O)$, which requires $r+s$ independent samples from $c_n^2$, we uniformly sample (with replacement) $r+s$ times from the $M$ samples we obtained with the single Markov chain. This bootstrapping technique allows us to avoid running $r+s$ independent Markov chains, making the computation significantly faster, but it works only if the $M$ samples faithfully reconstruct the original probability distribution. 

The number of samples needed to approximate the distribution clearly depends on the size of the system and usually increases exponentially with it. In order to estimate the value of $M$ for the sizes we are interested in (up to $L = 50$), we compare the sampled distribution with $M$ samples to the real one (for small system sizes where we can compute the probability density of the entire target distribution), computing the distance between the two. The distance we use is called \textit{total variation distance} and is simply defined as
\begin{equation}
    \mathrm{TV}(p, q) = \frac{1}{2} \sum_i |p_i - q_i|
\end{equation}
where $p_i$ and $q_i$ are the two discrete probability distribution functions. This probability distance is of easy interpretation, as it corresponds to the maximum difference in probability assigned to any event and, it is bounded between $0$ and $1$ and a threshold has a clear interpretation: $\mathrm{TV} = 0.1$ means "the distributions disagree on at most $10\%$ of the probability mass".

\begin{figure}[h]
    \centering
\includegraphics[width=0.5\linewidth]{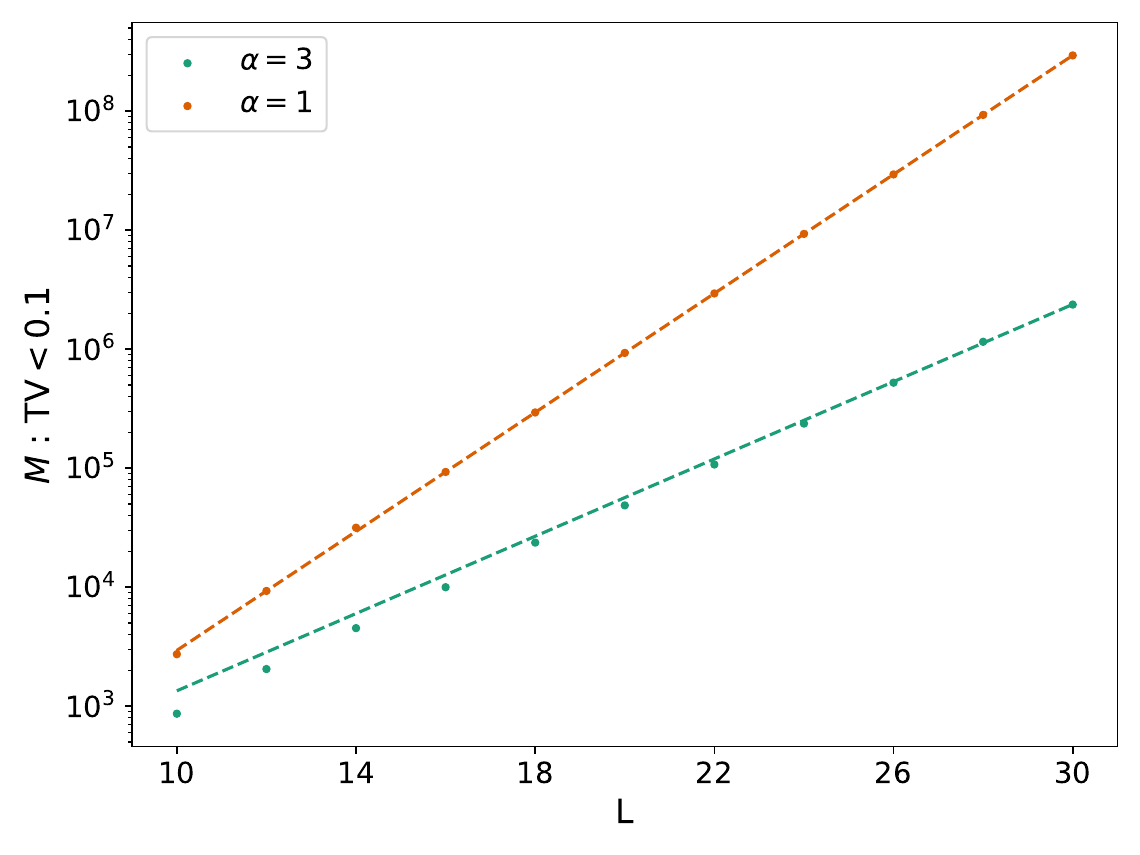}
    \caption{Number of the Markov chain samples of the JG distribution, for different values of $\alpha$, required to get the total variation distance under the $0.1$ threshold. The dashed lines represent an exponential fit $a e^{bL}$ on the data. For $\alpha=3$, we get $a=32, b=0.374$, and for $\alpha=1$, we get $a=9, b=0.575$.}
    \label{fig:mc_samples}
\end{figure}

$\mathrm{TV}= 0.1$ happens to be the threshold we chose by testing on small system sizes, as we see that it allows us to reach a maximum relative error (compared to the exact values obtained via ED) on the estimations of $F_n$ of Eq.~\ref{eq:bounds1} of a maximum of around $1\%$. Then to determine the number of samples we need for $L = 50$, by studying between $L=10$ and $L=30$ how many samples are required to be under the $\mathrm{TV} = 0.1$ threshold, fitting the results with an exponential function $n_\text{req}(L) = a \exp(b\, L)$ which we use to extrapolate the value of $M$ for $L = 50$. The results for $\alpha = 1$ and $\alpha = 3$ are reported in Fig.~\ref{fig:mc_samples} (for the other values of $\alpha$ considered in this work, the distribution is much simpler and many fewer samples are required). For $\alpha = 1$, where the exponential increase in the number of samples required is more pronounced, and the number of samples required for $L=50$ that the exponential fit predicts is $M \sim 10^{13}$.

For every system size $L$, we drew $M = 10^9$ samples. Based on the previous 
analysis, this sample count should only be potentially insufficient in the 
$\alpha = 1$ regime. Nevertheless, we remain confident that it is adequate even 
in this case, as the total variation ($\mathrm{TV}$) threshold below which the 
bounds carry a $1\%$ error grows with system size. Furthermore, results obtained 
for larger sizes are consistently compatible with known reference quantities. 
Regarding the Markov chain sampling procedure, we retained one sample every $L$ 
steps to mitigate the autocorrelation; for the same reason, all observables were 
estimated via block averaging, and the statistical uncertainty on the mean is 
derived from the standard deviation across multiple independent estimates of 
that mean.

\section{Local dephasing}\label{app:dephasing}

The first source of noise that we consider is the individual dephasing resulting from coupling of each hard-core boson to an independent noise
\begin{equation}\label{eq:dephasing_channel}
    \rho_0 \rightarrow \rho_p=\prod_j \mathcal{E}^Z_j\left[\rho_0\right] \quad \text{with} \quad \mathcal{E}^Z_j\left[\rho_0\right]=(1-p) \rho_0+p Z_j \rho_0 Z_j.
\end{equation}

\subsection{Analytical expression for the mixed state}\label{app:loc_deph_analytical}

We can write the pure state $\rho_0$ as
\begin{equation}\label{eq:density_matrix}
    \rho_0 = \ket{\psi}\bra{\psi} = \sum_{n n'} c_n c_{n'} \ket{n}\bra{n'}
\end{equation}
where $c_n$ corresponds to $\psi_\alpha(\{n\})$ of Eq.~\ref{eq:wf}, and applying the quantum channel~\ref{eq:dephasing_channel} the dephased mixed state becomes
\begin{equation}
    \rho_p = \prod_j \mathcal{E}^p_j\left[\rho_0\right] = \sum_{s=0}^L \sum_{|q| = s} p^s (1-p)^{L-s} Z_q \rho_0 Z_q
\end{equation}
where we defined $Z_q = \prod_{i \in q} Z_i$ and $q$ is the set size $s$ of the indices $j = 1,..,L$. We note that $Z_q\ket{n} = (-1)^{\sum_{i \in q}(1 - n_i)}\ket{n} = (-1)^{s - \sum_{i \in q}n_i}\ket{n}$, which results in
\begin{equation}
    Z_q \rho_0 Z_q = \sum_{n n'} (-1)^{\sum_{i \in q} n_i + n'_i} c_n c_{n'} \ket{n} \bra{n'}.
\end{equation}
By defining $n_i'' \equiv (n_i + n_i')\operatorname{mod} 2$ and $N'' \equiv \sum_{i=1}^L n''_i$ we have
\begin{equation}
    \sum_{|q| = s} Z_q \rho_0 Z_q = \sum_{nn'} \sum_{i=0}^{N''} (-1)^i \binom{N''}{i} \binom{L-N''}{s-i}c_n c_{n'} \ket{n}\bra{n'}.
\end{equation}
Summing over $s$ we obtain
\begin{equation}
    \sum_{nn'} \sum_{s=0}^L p^s (1-p)^{L-s} \sum_{i=0}^{N''} (-1)^i \binom{N''}{i} \binom{L-N''}{s-i}c_n c_{n'} \ket{n}\bra{n'}
\end{equation}
and we have that the pure state matrix element $c_nc_{n'}$ of the pure state are just multiplied by the factor
\begin{equation}
    f_{nn'}(p) \equiv \sum_{s=0}^L p^s (1-p)^{L-s} \sum_{i=0}^{N''} (-1)^i \binom{N''}{i} \binom{L-N''}{s-i}
\end{equation}
which involves only $n$ and $n'$. We can further simplify $f_{nn'}(p)$, by factoring out $(1-p)^L$, defining $x = \frac{p}{1-p}$ and making the change of variable $j=s-i$, with $j=0,1,\dots L$ we obtain
\begin{equation}
    f_{nn'}(p) = (1-p)^L \sum_{i=0}^{N''} (-1)^i x^i \binom{N''}{i} \sum_{j=0}^L \binom{L-N''}{j} x^j.
\end{equation}
Noting that $\binom{L-N''}{j} = 0$ for $j > L-N''$ the two sums are just the binomial factorizations
\begin{equation}
    (1-x)^{N''} = \sum_i (-1)^i \binom{N''}{i}x^i, \quad  (1+x)^{L-N''} = \sum_j \binom{L-N''}{j}x^j.
\end{equation}
Therefore, we have
\begin{equation}
f_{nn'}(p) = (1-p)^L \left(1-\frac{p}{1-p}\right)^{N''} \left(1+\frac{p}{1-p}\right)^{L-N''} = (1-2p)^{N''}.
\end{equation}
and we finally have the result of Eq.~\ref{eq:state_local_dephasing}.

\subsection{Monte Carlo estimator of the density matrix moments}

The main quantities of interest that we need to compute to obtain the bounds of Eq.~\ref{eq:bounds1} and Eq.~\ref{eq:bounds2} are $\operatorname{Tr}(\rho^r O \rho^s O)$. In the local dephasing case, where the channel is diagonal in the occupation-number basis, inserting the expression of the state in Eq.~\ref{eq:state_local_dephasing} in the traces gives
\begin{equation}
\begin{aligned}
    \operatorname{Tr}(\rho^r O \rho^s O) = \operatorname{Tr}\big(\sum_{\substack{n_1\dots n_{2r} \\\ m_1\dots m_{2s}}} & c_{n_1} c_{n_2}\dots c_{n_{2r}}c_{m_1}c_{m_2}\dots c_{m_{2s}} f_{n_1 n_2}  \dots f_{n_{2r-1} n_{2r}} f_{m_1 m_2} \dots f_{m_{2s-1} m_{2s}} \\ &\times  \ket{n_1} \braket{n_2|n_3}\dots\braket{n_{2r}|O|m_{1}}\braket{m_2|m_3}\braket{m_4|m_5}\dots\bra{m_{2s}}O\big).
\end{aligned}
\end{equation}
All the products $\braket{n_i|n_j}=\braket{m_i|m_j} =\delta_{ij}$ eliminate half of the terms in the sum and taking the trace gives us
\begin{equation}
\operatorname{Tr}\left(\rho_p^r O \rho_p^s O\right)
= \sum_{\substack{n_1\dots n_{r+1} \\ \ m_1\dots m_{s+1}}} c_{n_1}^2 c_{n_2}^2  \dots c_{n_r}^2 
c_{m_1}^2 c_{m_2}^2  \dots c_{m_s}^2
f_{n_1 n_2}  \dots f_{m_s m_{s+1}} \frac{c_{n_{r+1}} c_{m_{s+1}}}{c_{n_1} c_{m_1}}\langle n_{r+1}|O|m_1\rangle
\langle m_{s+1}|O|n_1\rangle 
\end{equation}
where we have defined $f_{n n'} = (1 - 2p)^{N''}$, coming from Eq.~\ref{eq:state_local_dephasing} and we relabeled the configurations $n_i$ and $m_i$. In this form, these traces appear as averages over the probability distributions $c_n^2$ of the pure JG state. By sampling these probability distributions using a Markov chain as described in App.~\ref{app:montecarlo}, the quantities can be estimated via Monte Carlo integration, with
\begin{equation}\label{eq:dephased_estimator}
    \left\langle \sum_{\substack{n_{r+1}\\m_{s+1}}} f_{n_1 n_2}  \dots f_{m_s m_{s+1}} \frac{c_{n_{r+1}} c_{m_{s+1}}}{c_{n_1} c_{m_1}}\langle n_{r+1}|O|m_1\rangle \langle m_{s+1}|O|n_1\rangle \right\rangle_{\substack{n_1\dots n_r \\ m_1 \dots m_s}}
\end{equation}
as an unbiased estimator. Although the summations on $n_{r+1}$ and $m_{s+1}$ are technically over all configuration space, making it unfeasible to compute, for the operators we are interested in, Eq.~\ref{eq:dephased_estimator} simplifies considerably.
Indeed, if $O$ is a linear combination of $Z_j$ operators, the brackets of the operators collapse the summations entirely. Instead for $O_X = \tfrac{1}{2}\sum_j (-1)^j X_j$ we note that $s$ can only be equal to $0$, since $c_n$ is different from $0$ only for configurations at half-filling and $X_j$, by flipping only one spin, cannot connect two half-filled configurations. Then in the case $s=0$ the estimator becomes
\begin{equation}
    \left\langle \sum_{n_{r+1}} \sum_{ij} f_{n_1 n_2} \dots f_{n_r n_{r+1}} \frac{c_{n_{r+1}}}{c_{n_1}}\langle n_{r+1}|X_iX_j|n_1\rangle \right\rangle_{n_1\dots n_r}
\end{equation}
where now the sum is equal to $0$ for all the configurations that are not at half-filling (otherwise $c_{n_{r+1}} = 0$ and that differs for more than two bit flips from $n_1$. These considerations severely restrict the terms in the sum in the estimator, making it feasible to compute for the system sizes considered in this work.

\subsection{QFI of the GHZ and Dicke states}\label{qfidicke}

Let us consider the GHZ state
\begin{equation}
    \ket{\text{GHZ}} = \frac{1}{\sqrt{2}}\left(\ket{\psi_\text{even}} + \ket{\psi_\text{odd}}\right),
\end{equation}
where $\ket{\psi_\text{even}} = \ket{0101\dots01}$ and $\ket{\psi_\text{odd}} = \ket{1010\dots10}$.
The corresponding density matrix, written in the occupation number basis as in Eq.~\ref{eq:density_matrix}, is just a 2 by 2 matrix with all elements equal to $1/2$. Using the results of Eq.~\ref{eq:state_local_dephasing}, we know that applying the local dephasing channel on the GHZ state has the effect of suppressing the coherences, i.e. the off diagonal terms of the density matrix, exponentially in $L$. Then for the GHZ state, the suppression factor is equal to $(1-2p)^L$ and the resulting density matrix has eigeinvalues $\lambda_{1,2} = (1 \pm (1-2p)^L) / 2$, with the corresponding eigenvectors being the symmetric and anti-symmetric GHZ state. As a consequence, it is trivial to see that, if the operator $O$ is diagonal in the occupation number basis, the QFI of the dephased state $\rho'$ is obtained from the QFI of the pure GHZ state $\rho$ simply as
\begin{equation}
F_Q[\rho', O] = (1-2p)^{2L} F_Q[\rho, O].
\end{equation}

In the strongly clustered limit $\alpha\ll -1$, the Jastrow--Gutzwiller wave function at half filling is well
approximated by the ``Dicke-like'' superposition of translated particle blocks,
\begin{equation}
	\ket{\Psi}=\frac{1}{\sqrt{L}}\sum_{i=1}^{L}\ket{\phi_i},
	\qquad
	\ket{\phi_i}=\ket{\cdots 0_{i-1}1_i1_{i+1}\cdots 1_{i+L/2-1}0_{i+L/2}\cdots},
\end{equation}
where $\{\ket{\phi_i}\}_{i=1}^L$ are orthonormal. We consider the operator $O_\star$ defined in Eq.~\ref{eq:obigstar}. Since $O_\star$ is diagonal in the Fock basis, it preserves the subspace
$V=\mathrm{span}\{\ket{\phi_i}\}_{i=1}^L$ and we can compute the QFI within this $L$-dimensional subspace. Again, the effect of the dephasing channel is to suppress the coherences between different block configurations and denoting $q\equiv 1-2p$, we have
\begin{equation}
	\rho_p
	=\frac{1}{L}\sum_{i,j=1}^L q^{N''(\phi_i,\phi_j)}\ket{\phi_i}\bra{\phi_j},
\end{equation}
where $N''(\phi_i,\phi_j)$ is the Hamming distance between the two bitstrings. For two compact blocks, shifting
the block by a distance $d_{ij}$ (on the ring) vacates $d_{ij}$ sites and occupies $d_{ij}$ new ones, hence
\begin{equation}
	N''(\phi_i,\phi_j)=2d_{ij},
	\qquad
	d_{ij}\equiv \min\!\big(|i-j|,\,L-|i-j|\big).
\end{equation}
The matrix $M_{ij}\equiv \bra{\phi_i}\rho_p\ket{\phi_j}=q^{2d_{ij}}/L$ is real, symmetric, and circulant, therefore it is diagonalized by the discrete Fourier modes
\begin{equation}
	\ket{m}=\frac{1}{\sqrt{L}}\sum_{j=1}^{L} e^{-2\pi i m j/L}\ket{\phi_j},
	\qquad m=0,1,\ldots,L-1,
\end{equation}
with eigenvalues equal to the discrete Fourier transform of the first row of the matrix, i.e.
\begin{equation}
	\mu_m=\frac{1}{L} \sum_{k=-L/2+1}^{L/2} q^{2k} e^{-i \theta_m k}= \frac{1}{L}\!\left[1+2\sum_{k=1}^{L/2-1} q^{2k}\cos(k\theta_m)+q^{L}(-1)^m\right],
	\qquad \theta_m=\frac{2\pi m}{L}.
	\label{eq:dicke_lambda_m}
\end{equation}
The remaining sum is geometric and combining the real parts we get
\begin{equation}\label{eq:dicke_mu_closed}
    \mu_m=\frac{1-q^4}{L\left(1+q^4-2 q^2 \cos \theta_k\right)}\left[1-(-1)^k q^L\right]
\end{equation}
Since $O_\star\ket{\phi_i}=O_i\ket{\phi_i}$, in the $\{\ket{m}\}$ the matrix elements are
\begin{equation}
	\braket{m|O_\star|m'}=\tilde O_{m-m'\,(\mathrm{mod}\ L)},
	\qquad
	\tilde O_k\equiv \frac{1}{L}\sum_{j=1}^{L}O_j\,e^{2\pi i k j/L}.
\end{equation}
Substituting into $F_Q=2 \sum_{k k^{\prime}}\left(\mu_k-\mu_{k^{\prime}}\right)^2 /\left(\mu_k+\mu_{k^{\prime}}\right)\left|\langle k| O_{\star}| k^{\prime}\right\rangle\left.\right|^2$ we obtain
\begin{equation}\label{eq:dicke_qfi_factorized}
    F_Q\left[\rho_p, O_{\star}\right]=2 \sum_{q=1}^{L-1}|\widetilde{O}_q|^2 G_q(L, p), \quad G_q(L, p) \equiv \sum_{k=0}^{L-1} \frac{\left(\mu_k-\mu_{k-q}\right)^2}{\mu_k+\mu_{k-q}} .
\end{equation}
In the thermodynamic limit at fixed $p\in(0,1/2)$, we first note that the factor $q^L$ in Eq.~\eqref{eq:dicke_mu_closed} is exponentially small in $L$ and can be dropped, so that
\begin{equation}
\mu_m \simeq \frac{1}{L} f(\theta_m),
\qquad
f(\theta)\equiv \frac{1-q^4}{1-2q^2\cos\theta+q^4}.
\label{eq:dicke_f_theta}
\end{equation}
Using Eq.~\eqref{eq:dicke_f_theta} in $G_r(L,p)$ yields
\begin{equation}
G_r(L,p)\simeq \sum_{m=0}^{L-1}\frac{\left[f(\theta_m)-f(\theta_{m-r})\right]^2}{f(\theta_m)+f(\theta_{m-r})}\,\frac{1}{L}.
\label{eq:G_sum_prefactor}
\end{equation}
For fixed integer $r$ and large $L$, we have $\theta_{m-r}=\theta_m-\delta$ with $\delta\equiv 2\pi r/L\ll 1$, so we expand
$f(\theta_m-\delta)=f(\theta_m)-\delta f'(\theta_m)+O(\delta^2)$, and therefore
\begin{equation}
\frac{\left[f(\theta_m)-f(\theta_{m-r})\right]^2}{f(\theta_m)+f(\theta_{m-r})}
=\frac{\delta^2 f'(\theta_m)^2}{2f(\theta_m)}+O(\delta^3).
\label{eq:small_delta_ratio}
\end{equation}
Replacing the Riemann sum by an integral, $\frac{1}{L}\sum_{m=0}^{L-1}(\cdots)\to\frac{1}{2\pi}\int_0^{2\pi}d\theta(\cdots)$, we obtain
\begin{equation}
G_r(L,p)\simeq \frac{\delta^2}{2}\,\frac{L}{2\pi}\int_0^{2\pi} d\theta\,\frac{f'(\theta)^2}{2f(\theta)}
=\frac{(2\pi r/L)^2}{4}\,\frac{L}{2\pi}\int_0^{2\pi} d\theta\,\frac{f'(\theta)^2}{f(\theta)}.
\label{eq:G_asymptotic}
\end{equation}
For the specific choice of $O_\star$, in the limit of large-$L$ we find
\begin{equation}
|\tilde O_r|^2 = \frac{1}{L^2\sin^4(\pi r/L)}\quad (r\ \text{odd}),\qquad |\tilde O_r|^2=0\quad (r\ \text{even}),
\label{eq:Otilde_triangular}
\end{equation}
so that for $r\ll L$ we can use $\sin(\pi r/L)\simeq \pi r/L$, giving $|\tilde O_r|^2\simeq L^2/(\pi^4 r^4)$ (odd $r$).
Inserting Eqs.~\eqref{eq:G_asymptotic} and \eqref{eq:Otilde_triangular} into Eq.~\eqref{eq:dicke_qfi_factorized}, the factors of $r$ combine as $|\tilde O_r|^2 G_r\sim (1/r^4)\times r^2\sim 1/r^2$, and the sum over odd $r$ converges to a constant:
\begin{equation}
F_Q[\rho_p,O_\star]
\simeq
2\sum_{\substack{r=1\\ r\ \mathrm{odd}}}^{\infty}
\left(\frac{L^2}{\pi^4 r^4}\right)
\left[
\frac{(2\pi r/L)^2}{4}\,\frac{L}{2\pi}\int_0^{2\pi} d\theta\,\frac{f'(\theta)^2}{f(\theta)}
\right]
=
\frac{1}{\pi^2}\left(\sum_{\substack{r=1\\ r\ \mathrm{odd}}}^{\infty}\frac{1}{r^2}\right)
\left(\frac{1}{2\pi}\int_0^{2\pi} d\theta\,\frac{f'(\theta)^2}{f(\theta)}\right).
\label{eq:FQ_constant_reduction}
\end{equation}
Using $\sum_{r\ \mathrm{odd}} r^{-2}=\pi^2/8$ and evaluating the remaining integral with $f(\theta)$ from Eq.~\eqref{eq:dicke_f_theta} (a straightforward rational integral after setting $z=e^{i\theta}$), one finds
\begin{equation}
\frac{1}{2\pi}\int_0^{2\pi} d\theta\,\frac{f'(\theta)^2}{f(\theta)}=\frac{64\,q^4}{\left(1-q^4\right)^2}.
\label{eq:integral_result}
\end{equation}
Combining Eqs.~\eqref{eq:FQ_constant_reduction} and \eqref{eq:integral_result} yields the thermodynamic-limit saturation value
\begin{equation}
F_Q[\rho_p,O_\star]\ \xrightarrow[L\to\infty]{}\ \frac{8\,q^4}{(1-q^4)^2}
=\frac{8(1-2p)^4}{\left[1-(1-2p)^4\right]^2}.
\label{eq:dicke_qfi_thermo_final}
\end{equation}
For small $p$, $1-(1-2p)^4\simeq 8p$ and therefore $F_Q\simeq 1/(8p^2)$.

\subsection{Analytical bounds on the QFI in the critical regime}\label{app:sarabound}

We can compute another bound to the QFI using the technique proposed in \textcite{Sara}. Considering the local dephasing channel, the $Z$ parity symmetry is conserved, while the $X$ and $Y$ parity symmetries are not, hence we have two cases: for $\alpha \in [0, 2]$ where the optimal imprinter is the $X$ operator, which anti-commutes with the generator of the symmetry $Z$, allows us to compute the QFI directly, starting from the one of the pure state as~\cite{Sara}
\begin{equation}
    F_Q[\rho']=4(1-2 p)^2\left\langle O^2\right\rangle_\rho +4 p(1-p) L.
\end{equation}
For $\alpha \in (2, 4]$ the $Z$ operator is the optimal imprinter, and we can bound the scaling of the QFI by choosing an operator $S_0$ to compute 
\begin{equation}
    \delta \theta = \left| \frac{\sqrt{\operatorname{Var}_{\rho} S_\theta}}{\partial_\theta\langle S_\theta\rangle_{\rho}}\right|
\end{equation}
where $S_\theta = e^{-i \sum_j (-1)^j Z_j} S_0 e^{i \sum_j (-1)^j Z_j}$. In fact, the quantum Cramér–Rao bound tells us that the best precision on the estimation of the parameter $\theta$ is bounded by $\delta \theta \ge \frac{1}{\sqrt{F_Q[\rho, O]}}$ and it can be shown that, if we choose $S_0$ so that it anticommutes with $O$, the scaling of $\delta \theta$ gives us a lower bound to the scaling of the QFI itself.

We choose $S_0 = \sum_i X_i X_{i+1}$ and obtain
\begin{equation}
    S_\theta = \sum_i \cos^2(2\theta) X_iX_{i+1} + (-1)^i \frac{\sin(4 \theta)}{2} \left(Y_i X_{i+1} - X_i Y_{i+1} \right) - \sin^2 (2 \theta) Y_i Y_{i+1}.
\end{equation}
For now we consider the pure state and we take the expectation value. By the symmetries of the state, for which $\braket{X_i X_{i+1}} = \braket{Y_i Y_{i+1}}$ and $\braket{Y_iX_{i+1}} = \braket{X_i Y_{i+1}}$, we have that
\begin{equation}
    \braket{S_\theta} = L \cos{4\theta} \braket{X_i X_{i+1}}, \quad \partial_\theta \braket{S_\theta} = - 4 L \sin{4\theta} \braket{X_i X_{i+1}}.
\end{equation}
We conclude that the optimal value of $\theta$ to measure at is around $\theta = \pi/8$, for which the expectation value is equal to zero and then the variance appearing in the formula for $\delta \theta$ will only be equal to the expectation value of $S_\theta^2$. By similar calculations, setting $\theta = \pi / 8$, we have
\begin{equation}
\begin{aligned}
    \braket{S_\theta^2} = \sum_{ij}& \frac{1}{2}\braket{X_i X_{i+1} X_j X_{j+1}} - \frac{1}{2} \braket{X_i X_{i+1} Y_j Y_{j+1}} +
\\ + &\frac{1}{2} ((-1)^{i+j} \braket{X_i Y_{i+1} X_j Y_{j+1}} + (-1)^{i+j+1} \braket{X_i Y_{i+1} Y_j X_{j+1}}).
\end{aligned}
\end{equation}

To estimate the bound on the mixed state we can use the results derived so far for the pure state because we can write $\braket{S_\theta} = \operatorname{Tr}\left\{\prod_j \mathcal{E}_j\left[\rho_0\right] S_\theta\right\}=\operatorname{Tr}\left\{\rho_0 \prod_j \mathcal{E}_j^*\left[S_\theta\right]\right\}$ and $ \left\langle S_\theta^2\right\rangle=\operatorname{Tr}\left\{\prod_j \mathcal{E}_j\left[\rho_0\right] S_\theta^2\right\}=\operatorname{Tr}\left\{\rho_0 \prod_j \mathcal{E}_j^*\left[S_\theta^2\right]\right\}$. 
Here we have used the fact that, since correlators are linear objects in $\rho$, the action of $\mathcal{E} [\cdot] = \prod_j \mathcal{E}_j [\cdot]$ on $\rho$ can be equivalently computed as the action of its conjugate channel $\mathcal{E}^* [\cdot]$ (which in this case is the same as with $\mathcal{E} [\cdot]$). We note that
\begin{equation}
    \mathcal{E}_j[X_i X_{i+1}] = (1 - 2p) (\delta_{ij} + \delta_{i+1,j}) X_i X_{i+1} \quad \text{and} \quad \prod_j \mathcal{E}_j [X_i X_{i+1}] = (1 - 2p)^2 X_i X_{i+1}
\end{equation}
and, accounting for all the terms of $S_\theta$, result in
\begin{equation}
    \prod_j \mathcal{E}_j [S_\theta] = (1-2p)^2 S_\theta.
\end{equation}
Following a similar procedure, we obtain
\begin{equation}
\prod_k \mathcal{E}_k\!\left[S_\theta^2\right]
= (1-2p)^4\, S_\theta^2
+ 2\,\bigl[(1-2p)^2 - (1-2p)^4\bigr]\, S_\theta^{(2)}
+ L\,\bigl[1 - (1-2p)^4\bigr]\, I, 
\label{eq:channel-S2-final}
\end{equation}
where we defined
\begin{equation}
S_\theta^{(2)} \;\equiv\; \sum_i\!\Big[\cos^2(2\theta)\, X_i X_{i+2}
+ (-1)^i\tfrac{\sin(4\theta)}{2}\!\left(Y_i X_{i+2} + X_i Y_{i+2}\right)
+ \sin^2(2\theta)\, Y_i Y_{i+2}\Big].
\end{equation}

Now, we observe that all the terms of the kind $\sum_i\braket{X_i X_{i+1}}$ and $\sum_i\braket{Y_i Y_{i+2}}$ will scale like $L$, therefore we only need to determine the scaling of $\braket{S_\theta^2}$ to finally determine the scaling of $\delta\theta$. 

Using the Luttinger liquid theory that allowed us to obtain the correlation functions and the variance of the operators as detailed in~\ref{app:luttinger}, it is easy to show that $\braket{S_\theta^2} \sim 1/L^\alpha$. Therefore, since we consider the interval $\alpha \in (2, 4]$, the variance of $S_\theta$ over the mixed state scales linearly in $L$ and $\delta \theta \sim 1 / \sqrt{L}$. As a result, the QFI is bounded to scale at most linearly for $\alpha=3$, which seems to be the case if we consider the exact results for small systems we plot in Fig.~\ref{fig:dephasing}.

\section{Local amplitude damping}
\label{app:damping}

In the amplitude damping case, the channel with strength $p$ acts locally on the spin $j$ via the Kraus operators
\begin{equation}
    K_j^{(0)} = |0\rangle\langle 0|_j + \sqrt{1-p}\,|1\rangle\langle 1|_j,
    \qquad
    K_j^{(1)} = \sqrt{p}\,|0\rangle\langle 1|_j,
\end{equation}
and the total effect of the channel on the state is given by the tensor product over all $L$ sites
\begin{equation}
    \mathcal{E}_p(\,\cdot\,)
    = \sum_{q \subseteq \{1,\ldots,L\}}
    M_q\,(\,\cdot\,)\,M_q^\dagger,
    \qquad
    M_q = \prod_{j \in q} K_j^{(1)}\prod_{j \notin q} K_j^{(0)},
\end{equation}
where the sum runs over all $2^L$ subsets $q$ of lattice sites.

\subsection{Analytical expression for the mixed state}

For a fixed Kraus branch $q$, the action on the basis state $\ket{n}$
\begin{equation}
    M_q |n\rangle =
    \begin{cases}
        p^{|q|/2}(1-p)^{(N-|q|)/2}\,\ket{m}
        & \text{if } q \subseteq n, \\[4pt]
        0 & \text{otherwise,}
    \end{cases}
\end{equation}
where $m$ denotes the configuration $n$ with all sites in $q$ vacated, $|q|$ is the cardinality of $q$. From now on we will refer to $n$ as the set of indices of the occupied sites in the configuration $n$, such that $n = m \cup n$ and $N = L/2 = |n|$. Applying $\mathcal{E}_p$ to the pure state $\rho_0 = |\psi_\alpha\rangle\langle\psi_\alpha| = \sum_{nn'} c_n c_{n'} \ket{n}\bra{n'}$ (where, as for the dephasing case, $c_n$ corresponds to $\psi_\alpha(\{n\})$ of Eq.~\ref{eq:wf}) gives
\begin{align}
    \rho_p
    &= \mathcal{E}_p\!\left(|\psi_\alpha\rangle\langle\psi_\alpha|\right)
    = \sum_{q \subseteq \{1,\ldots,L\}}
      M_q\,|\psi_\alpha\rangle\langle\psi_\alpha|\,M_q^\dagger \notag \\
    &= \sum_{q \subseteq \{1,\ldots,L\}}\,
      p^{|q|}(1-p)^{N-|q|}
      \left(\sum_{\substack{m:\, n = m\cup q}}
      c_n\,|m\rangle\right)
      \left(\sum_{m':\, n' = m'\cup q}
      c_{n'}\,\langle m'|\right).
\end{align}
By introducing the (unnormalized) conditional states
\begin{equation}\label{eq:psi_q}
    |\tilde{\psi}_q\rangle
    \equiv \sum_{m}
    c_{m\cup q}\,|m\rangle.
\end{equation}
where we omit the constraint $m\cup q=n$, since $c_n = 0$ for $|n| \neq N$ and we obtain for the mixed state
\begin{equation}\label{eq:rho_p}
    \rho_p = \sum_{q \subseteq \{1,\ldots,L\}}
    p^{|q|}(1-p)^{N-|q|}
    \,|\tilde{\psi}_q\rangle\langle\tilde{\psi}_q|.
\end{equation}

\subsection{Monte Carlo estimator of the density matrix moments}

Once more, in the computation of the QFI bounds, we are interested in calculating, through Monte Carlo integration, the moments $\operatorname{Tr}(\rho^r O \rho^s O)$. Similarly to the calculations for the dephased state, we set $P = r + s$ and we have
\begin{align}\label{eq:trace_expand}
    \operatorname{Tr}\!\left(\rho_p^r O \rho_p^s O\right)
    &= \sum_{q_1,\ldots,q_p}
    \left[\prod_{a=1}^{P} p^{|q_a|}(1-p)^{N-|q_a|}\right]
    \operatorname{Tr}\!\left(
      |\tilde\psi_{q_1}\rangle\langle\tilde\psi_{q_1}|
      \cdots
      |\tilde\psi_{q_{r}}\rangle\langle\tilde\psi_{q_{r}}|
      O
      |\tilde\psi_{q_{r+1}}\rangle\langle\tilde\psi_{q_{r+1}}|
      \cdots
      |\tilde\psi_{q_{P}}\rangle\langle\tilde\psi_{q_{P}}|
      O
    \right).
\end{align}
We start from the case where the operator is diagonal in the occupation number basis, i.e. $O|m\rangle = \mathcal{O}(m)|m\rangle$ and for the overlaps we have
\begin{equation}\label{eq:overlaps}
    \langle\tilde\psi_{q_a}|\tilde\psi_{q_b}\rangle
    = \sum_{m} c_{m\cup q_a}\, c_{m\cup q_b}, \quad \text{and} \quad \langle\tilde\psi_{q_a}|O|\tilde\psi_{q_b}\rangle 
    = \sum_{m} c_{m\cup q_a}\, c_{m\cup q_b}\mathcal{O}(m),
\end{equation}
where the sum is over configurations $m$ compatible with both $q_a$ and $q_b$ (i.e. $|m|=N-|q_a|=N-|q_b|$, requiring $|q_a|=|q_b|$). This immediately shows that \emph{only terms with all $|q_a|$ equal
contribute}: setting $|q_a| = k$ for all $a$, the sum collapses to
\begin{equation}\label{eq:trace_k}
    \operatorname{Tr}\!\left(\rho_p^r O \rho_p^s O\right)
    = \sum_{k=0}^{N}
    \left[p^k(1-p)^{N-k}\right]^P
    T_P^{(r,s)}(k),
\end{equation}
with
\begin{align}\label{eq:Trs_k}
    T_P^{(r,s)}(k)
    &=
    \sum_{\substack{q_1,\ldots,q_{P} \\ |q_a|=k\;\forall a}}
    \sum_{m_1,\ldots,m_{P}}
    \left[\prod_{a=1}^{P}
      c_{m_a \cup q_a}\,
      c_{m_a \cup q_{a+1}}
    \right]
    \mathcal{O}(m_{r+1})\,\mathcal{O}(m_1).
\end{align}
where we set $q_{P+1} = q_1$. Then for every link $a$, we introduce the full $N$ particle configurations
\begin{equation}
    n_a = m_a \cup q_a, \qquad a = 1,\ldots,P,
\end{equation}
and the product $c_{m_a\cup q_a} c_{m_a\cup q_{a+1}}$ can be written as
\begin{equation}
    c_{n_a} \frac{c_{m_a\cup q_{a+1}}}{c_{n_a}} c_{n_a}
    = c_{n_a}^2
    R\!\left(n_a,\, q_a,\, q_{a+1}\right),
\end{equation}
where
\begin{equation}\label{eq:ratio_def}
    R(n_a,\, q_a,\, q_{a+1}) \equiv  \frac{c_{m_a\cup q_{a+1}}}{c_n}
\end{equation}
is the JG amplitude ratio for removing the particles in $q_R$ from $n$ and inserting the particles in $q_A$. Substituting into \eqref{eq:Trs_k} and summing over each $q_a$ uniformly over all $k$-subsets of $n_a$, we obtain
\begin{equation}
    T_P^{(r,s)}(k) = \binom{N}{k}^P
    \sum_{n_1,\ldots,n_{P}}
    \left[\prod_{a=1}^{P} c_{n_a}^2\right]
    \notag
    \sum_{\substack{q_a :\, q_a \subseteq n_a\\ |q_a|=k }} \left[
    \prod_{a=1}^{P}
    R\left(n_{a},\;
    q_{a},\; q_{a+1}\right)
    \mathcal{O}(m_{r+1})\,\mathcal{O}(m_1)
    \right].
\end{equation}
Recognizing $c_n^2$ as the probability weight in the JG distribution and combining with \eqref{eq:trace_k}, we arrive at
\begin{equation}\label{eq:final_estimator}
    \operatorname{Tr}\!\left(\rho_p^r O \rho_p^s O\right)
    = Z_w^{(P)}\,
    \mathbb{E}_{n_a \sim |c^{(\alpha)}|^2}\,
    \mathbb{E}_{k \sim w_k^{(P)}}\,
    \mathbb{E}_{q_a \sim \mathrm{Unif}_k(n_a)}
    \left[
    \prod_{a=1}^{P}
    R\left(n_{a},\;
    q_{a},\; q_{a+1}\right)
    \mathcal{O}(m_{r+1})\,\mathcal{O}(m_1)
    \right]
\end{equation}
where the $k$-importance weights and their normalization are
\begin{equation}\label{eq:Zw}
    w_k^{(P)} = \frac{1}{Z_w^{(P)}}\binom{N}{k}^{\!P}
    p^{kP}(1-p)^{(N-k)P},
    \qquad
    Z_w^{(P)} = \sum_{k=0}^{N}\binom{N}{k}^{\!P}
    p^{kP}(1-p)^{(N-k)P}.
\end{equation}
Equation~\eqref{eq:final_estimator} provides an unbiased estimator:
for each Monte Carlo sample one draws $k\sim w_k^{(P)}$, $n_a \sim |c_n^{(\alpha)}|^2$, and $q_a \sim \mathrm{Unif}_k(\operatorname{supp}(n_a))$ independently for
each link $a$, and evaluates the ring product of amplitude ratios
\eqref{eq:ratio_def} together with the operator eigenvalues at the
post-damping configurations $m_a = n_a\setminus q_a$.
The estimator is then multiplied by the correction factor $Z_w^{(P)}$.

We now consider how to extend the Monte Carlo estimator to non-diagonal operators, focusing on the staggered transverse operator $O_X = \tfrac{1}{2}\sum_j (-1)^j X_j$, of relevance for this work. The main difference with respect to the diagonal $Z$-type operators considered above is that $O_X$ does not preserve the particle-number sector. Therefore, the operator insertions can connect different amplitude-damping branches with different numbers of lost particles. Then we have that the terms in Eq.~\ref{eq:overlaps} with $ O =  O_X$ become
 \begin{equation}\label{eq:ringX}
	c_{n_a}^2 R_X(n_a,q_a,q_{a+1})
	=
	c_{n_a}^2\sum_{j=1}^L
	(-1)^j
	\frac{
		c_{m_a^j\cup q_{a+1}}
	}{
		c_{n_a}
	}, \quad
    c_{n_a}^2 R_{X^2}(n_a,q_a,q_{a+1})
	=
	c_{n_a}^2 \sum_{j,l=1}^L
	(-1)^{j+l}
	\frac{
		c_{m_a^{jl}\cup q_{a+1}}
	}{
		c_{n_a}
	}.
\end{equation}
where $m_a^{j}$ denotes the configuration obtained from $m$ by flipping the occupation of site $j$ and $m_a^{jl}$ by flipping the occupation of site $j$ and $l$. The double flip term is relevant in the case $s=0$ in $\operatorname{Tr}(\rho^rO\rho^s O)$. These contributions are nonzero respectively only when
\begin{equation}\label{eq:ring_condX}
	\big||q_{a+1}| - |q_a|\big| = 1, \quad |q_{a+1}| - |q_a| \in \{-2,0,2\}.
\end{equation}
Then to obtain the unbiased estimator we can use the same procedure outlined before for the diagonal operator, but not all the $k$-subsets will be of the same cardinality. This means that we have to sample more than one value of $k$ and adjust the resulting $k$-importance sampling weights and their normalization. This consideration and the sums over the indices in Eq.~\ref{eq:ringX} make the evaluation of the moments $\operatorname{Tr}(\rho^rO\rho^sO)$ much more computationally expensive then the diagonal case.

\subsection{Analytical result for the QFI of the GHZ state}

Let us consider a system of $L$ qubits prepared in the GHZ state.
\begin{equation}
    \ket{\text{GHZ}} = \frac{1}{\sqrt{2}}\left(\ket{\psi_\text{even}} + \ket{\psi_\text{odd}}\right),
\end{equation}
where $\ket{\psi_\text{even}} = \ket{0101\dots01}$ and $\ket{\psi_\text{odd}} = \ket{1010\dots10}$.
For each branch $\ket{\psi_\sigma}$ (with $\sigma \in \{\text{even}, \text{odd}\}$), we define the following sets:
\begin{itemize}
    \item $n_\sigma$: the set of occupied sites without noise ($|n_\sigma| = L/2$),
    \item $q_\sigma$: the set of particles lost due to damping (sites where $K^{(1)}$ acts).
    \item $m_\sigma = n_\sigma \setminus q_\sigma$: the set of remaining occupied sites after damping.
\end{itemize}
Note that $n_\text{even} \cap n_\text{odd} = \emptyset$ and $|n_\text{even}| = |n_\text{odd}| = L/2$.

The action of the local damping channel on $\ket{\text{GHZ}}$ can be written as follows. Since $K_j^{(1)}\ket{0}_j = 0$, we need $q_\sigma \subseteq n_\sigma$ for the branch $\ket{\psi_\sigma}$ to survive. Given that $n_\text{even} \cap n_\text{odd} = \emptyset$, there are two cases:
 
\begin{enumerate}
    \item \textbf{$q_\sigma = \emptyset$} ($m_\sigma = n_\sigma$ for both branches): Both branches survive.
    \begin{equation}
        M_\emptyset \ket{\text{GHZ}} = \frac{(1-p)^{L/2}}{\sqrt{2}}\left(\ket{\psi_\text{even}} + \ket{\psi_\text{odd}}\right).
    \end{equation}
    Each branch has $|n_\sigma| = L/2$ occupied sites, each contributing a factor $\sqrt{1-p}$ from $K_j^{(0)}$.
 
    \item \textbf{$q_\sigma \neq \emptyset$, $q_\sigma \subseteq n_\sigma$} (with $|q_\sigma| = k$): Only the branch $\ket{\psi_\sigma}$ survives. The remaining sites are $m_\sigma = n_\sigma \setminus q_\sigma$, with $|m_\sigma| = L/2 - k$.
    \begin{equation}
        M_{q_\sigma}(\ket{\text{GHZ}}) = \sum_{m_\sigma} \frac{1}{\sqrt{2}}\,p^{k/2}\,(1-p)^{L/2-k}\,\ket{m_\sigma},
    \end{equation}
    where the sum runs over of states $\ket{m_\sigma}$ with $|m_\sigma|$ particles on the $\sigma = \text{even}, \text{odd}$ sites.
\end{enumerate}

Eventually, the output state $\rho' = \mathcal{E}_p(\ket{\text{GHZ}}\!\bra{\text{GHZ}})$ can be written as a sum over all possible $m_\sigma$:
\begin{equation}\label{eq:rho_prime}
    \rho' = (1-p)^{L/2}\ket{\text{GHZ}}\bra{\text{GHZ}} + \sum_{\sigma=\text{even},\\ \text{odd}}\sum_{m_\sigma} \frac{p^{|q_\sigma|}(1-p)^{|m_\sigma|}}{2}\ket{m_\sigma}\bra{m_\sigma}
\end{equation}
where $|m_\sigma|<\frac{L}{2}$ and $|q_\sigma| = \frac{L}{2}-|m_\sigma|$.

We observe that $\rho'$ is diagonal in the computational basis, except for the $2\times 2$ coherent block that corresponds to the $\ket{\text{GHZ}}$ state, weighted by $(1-p)^\frac{L}{2}$. Considering the QFI in Eq.~\ref{eq:qfi} with the staggered magnetization operator $O_Z = \tfrac{1}{2} \sum_{j=1}^L (-1)^jZ_j$, we note that since $O$ is diagonal in the computational basis, all pairs of diagonal eigenstates $\ket{m_\sigma}$ give $\braket{m_\sigma|O|m'_{\sigma'}} = 0$ for $m_\sigma \neq m'_{\sigma'}$. The only non-vanishing contribution comes from the $\ket{\text{GHZ}}$ state. We obtain
\begin{equation}
    F_Q[\rho', O] = (1-p)^{L/2}\,F_Q[\rho, O],
\end{equation}
where $F_Q[\rho, O] = L^2$ is the QFI of the pure GHZ state.

\end{document}